\documentclass[]{jfm}

\usepackage{graphicx}
\usepackage{newtxtext}
\usepackage{newtxmath}
\usepackage{natbib}

\usepackage{hyperref}
\usepackage{subcaption}
\usepackage{multirow}
\usepackage{longtable}
\hypersetup{
    colorlinks = true,
    urlcolor   = blue,
    citecolor  = black,
}

\newcommand{\RomanNumeralCaps}[1]
\title{Realizability-Informed Machine Learning for Turbulence Anisotropy Mappings}

\author{Ryley McConkey\aff{1,2}
  \corresp{\email{rmcconke@mit.edu}},
  Nikhila Kalia\aff{2},
  Eugene Yee\aff{2}
 \and Fue-Sang Lien\aff{2}}

\affiliation{\aff{1}Massachusetts Institute of Technology, 77 Massachusetts Avenue Cambridge, MA 02139-4307
\aff{2}200 University Ave W, Waterloo, ON N2L 3G1, Canada}

\begin{document}
\maketitle

\begin{abstract}
Within the context of machine learning-based closure mappings for RANS turbulence modelling, physical realizability is often enforced using ad-hoc postprocessing of the predicted anisotropy tensor. In this study, we address the realizability issue via a new physics-based loss function that penalizes non-realizable results during training, thereby embedding a preference for realizable predictions into the model. Additionally, we propose a new framework for data-driven turbulence modelling which retains the stability and conditioning of optimal eddy viscosity-based approaches while embedding equivariance. Several modifications to the tensor basis neural network to enhance training and testing stability are proposed. We demonstrate the conditioning, stability, and generalization of the new framework and model architecture on three flows: flow over a flat plate, flow over periodic hills, and flow through a square duct. The realizability-informed loss function is demonstrated to significantly increase the number of realizable predictions made by the model when generalizing to a new flow configuration. Altogether, the proposed framework enables the training of stable and equivariant anisotropy mappings, with more physically realizable predictions on new data. We make our code available for use and modification by others. Moreover, as part of this study, we explore the applicability of Kolmogorov-Arnold Networks (KAN) to turbulence modeling, assessing its potential to address non-linear mappings in the anisotropy tensor predictions and demonstrating promising results for the flat plate case.
\end{abstract}

\begin{keywords}
Authors should not enter keywords on the manuscript, as these must be chosen by the author during the online submission process and will then be added during the typesetting process (see \href{https://www.cambridge.org/core/journals/journal-of-fluid-mechanics/information/list-of-keywords}{Keyword PDF} for the full list).  Other classifications will be added at the same time.
\end{keywords}

It is prohibitively expensive to resolve all relevant scales of turbulence for industrially relevant flows. Even with increasing computational capacity, Kolmogorov microscale-resolving techniques such as Direct Numerical Simulation (DNS) will be out of reach for decades~\citep{CFD2030}. In order to enable the practical simulation of turbulent flows, a variety of techniques are currently in use, such as Reynolds-Averaged Navier Stokes (RANS), Detached Eddy Simulation (DES), and Large Eddy Simulation (LES). Each technique comes with its own advantages. The user's available computational resources are the primary consideration. RANS, a technique which models turbulence as a single-scale phenomenon, remains the most popular industrial technique due to the computational costs of scale resolving simulations~\citep{Witherden2017}.

In recent years, a new menu item for industrial turbulence modelling is emerging. Machine learning has been used to augment RANS, DES, LES, and DNS, with various objectives, such as accelerating simulation times~\citep{Kochkov2021}, infer optimal coefficient fields~\citep{Duraisamy2017_fieldinversion}, model calibration~\citep{Matai2019}, and turbulence model augmentation~\citep{Brunton2020,Duraisamy2019}. For LES, emerging research on training in-situ and within a differentiable solver offers the prospect of a generalizable subgrid model~\citep{List_Chen_Thuerey_2022, Sirignano_MacArt_2023}. It is clear that for LES, allowing the model to interact with a solver for an unsteady flow vastly improves the generalizability of the learned closure relationship~\citep{List_Chen_Thuerey_2022}. Within the context of steady-state RANS simulations, a promising new menu item is the ability to train a specialized turbulence model with a given industrial dataset. Significant attention has been given to the development of machine learning-augmented closure models. Specifically, flow-specific sensitization of the stable but often inaccurate linear eddy viscosity relationship via machine learning is an area of major interest. In their seminal work, \citet{Ling2016} proposed a neural network architecture based on a tensor basis expansion of the anisotropy tensor, referred to as a tensor basis neural network (TBNN). This architecture was extended to random forests by \citet{Kaandorp2018}, and \citet{Kaandorp2020}. The TBNN architecture has been used in several studies, such as by \citet{Song2019} and \citet{Zhang2019a}. Further modifications to the TBNN framework within the context of simple channel flows have been proposed by \citet{Cai2022,Cai2024}. Specifically, Cai et al. studied issues related to non-unique mappings between the closure term and input features for plane channel flow, an issue reported by \citet{Liu2021}. This issue primarily occurs when a small input feature set is used, such as the 5 invariants (several of which are zero for 2D flows) used in Ling et al.'s original TBNN~\citep{Ling2016}. Several strategies have been proposed to address the issue of a non-unique mapping, including incorporation of additional input features by \citet{Wu2018}, and ensembling through a divide and conquer approach by \citet{Man2023}. In the present investigation, we address the issue related to non-uniqueness of the mapping via use of a rich input feature set. In an effort to make these mappings more transparent, \citet{Mandler2023} analyzed predictions made via an anisotropy mapping using input feature importance metrics such as SHapley Additive exPlanations (SHAP) values. Interpretability analysis sheds light on which RANS features contain more information about different flow physics, which helps guide future input feature selection~\citep{Mandler2023}. TBNN-type models are not the only architecture in use---others include the eigenvalue reconstruction technique proposed by \citet{Wu2018,Wu2019a}, and the Reynolds force vector approach proposed by \citet{Cruz2019}, and further investigated by \citet{Brener2022} and \citet{Amarloo2022}.

%Specific Problem Considered
Here, we consider the specific problem of training data-driven anisotropy mappings between RANS and a higher fidelity closure term from LES or DNS. Several open questions remain in the area of machine learning anisotropy mappings for RANS, with perhaps the most popular question being whether a ``universal turbulence model" could be produced via machine learning~\citep{Duraisamy2021,Spalart2023}. However, when it comes to generalizability of these mappings, the no-free-lunch theorem is at play as demonstrated in \citet{McConkey2022b}. Accurate sensitization of the anisotropy mapping for a given flow comes at the cost of generalization to completely new flows. Nevertheless, numerous studies show that the sensitized mapping generalizes well within the same flow type~\citep{McConkey2022b,Wu2018,Kaandorp2020,Man2023}. While this lack of new flow generalizability means that a machine learning augmented turbulence model is off the menu for some applications, it remains on the menu for many industrial applications. For example, an industrial user with LES data can leverage these techniques to develop an augmented RANS turbulence model sensitized to a flow of interest. Our opinion is that this lack of generalizability is due to a lack of sufficient data. Despite numerous datasets for this purpose being available, we believe the entire space of possible mappings between RANS input features and higher fidelity closure terms is still sparsely covered by available data. For the foreseeable future, these mappings will therefore be limited to flow-specific sensitization.

%Contribution
While the generalizability question is perhaps the most popular one, several other key issues remain in training anisotropy mappings for RANS. How can predicted quantities be injected in a stable and well-conditioned manner? How can eddy viscosity-based approaches be united with tensor basis neural network-type approaches? How can we incorporate alternatives to the traditional fully connected layers in a tensor basis neural network? In this investigation, we address several of these remaining questions. We formulate an injection framework which unites tensor basis neural network (TBNN)-type model architectures with the more stable optimal eddy viscosity-based techniques. This unification enables stable use of the equivariance-enforcing TBNN within the momentum equation, without the use of stabilizing blending factors~\citep{Kaandorp2020}. The proposed framework also has the advantage of producing a well-conditioned solution without the use of an optimal eddy viscosity, an often-unstable quantity that is difficult to predict via a machine learning model. We also target the problem of producing realizable predictions via a TBNN-type architecture via a physics-based loss function penalty (viz., incorporation of a learning bias in the framework). To further improve the flexibility and representational capacity of TBNN-type models, we investigate the inclusion of a Kolmogorov-Arnold network (KAN) \citep{liu2024kan} into the framework to replace the multi-layer perception in the TBNN. We also further investigate the invariant input feature sets commonly used for TBNN architectures, and provide new insights as to which input features are appropriate for use in flows with certain zero gradient directions. We demonstrate good generalization performance of the ``realizability-informed" TBNN. We also demonstrate that the realizability-informed loss function greatly reduces non-realizable predictions when generalizing.

%we investigate the inclusion of the Kolmogorov-Arnold network (KAN) - add a reference (ziming liu paper) into the framework to replace the multi-layer perception in the TBNN. -- something like this
%explain what TBKAN is in the introduction 

%Differences in what we're doing
The present work unites TBNN-type frameworks (for example: \citet{Ling2016}, \citet{Kaandorp2018}, \citet{Kaandorp2020}, and \citet{Man2023}) with optimal eddy viscosity frameworks (e.g. \citet{Wu2018}, \citet{Brener2021}, and \citet{McConkey2022}). The advantage here is maintaining a stable injection environment \citep{Wu2019,Brener2021}, while also retaining the simplicity, elegance, and implicit equivariance of the TBNN architecture. Additionally, the present work is the first to implement a way to inform the TBNN of physical realizability during the training process. Whereas most existing techniques to enforce realizability of the neural network involve an ad-hoc post-processing step, our technique leverages physical realizability as an additional training target, thereby embedding an additional physics-based (learning) bias into the model. This idea extends the learning bias proposed by \citet{Riccius2023}. Lastly, despite widespread use of minimal integrity basis input features for 2D flows, and flows through a square duct, the present investigation is the first to systematically examine these input features for flow through a square duct. This examination leads to several unsuitable input features being identified, and a codebase for investigating other flows of interest.

%Roadmap
This manuscript is organized as follows. Section~\ref{sec:methodology_realizability} describes the novel techniques in detail, including the injection framework (\ref{sec:injection_ri}--\ref{sec:conditioning_ri}), realizability-informed loss function (\ref{sec:realizability_informed_training}), and improved TBNN architecture (\ref{sec:NN_architecture}). Details on the datasets, input features, hyperparameters, implementation, and code availability are given in Section~\ref{sec:machine_learning}. Two primary results are presented in Section~\ref{sec:results_ri}: generalization tests of the realizability-informed TBNN (\ref{sec:generalizability_tests}), and an examination of how realizability-informed training produces more realizable predictions when generalizing (\ref{sec:realizability_results}). Conclusions and future work are discussed in Section~\ref{sec:conclusion_ri}.

\section{Methodology}\label{sec:methodology_realizability}

\subsection{Injecting Reynolds stress tensor decompositions used in data-driven closure modelling frameworks}\label{sec:injection_ri}

%RANS Equation
The steady-state, Reynolds-Averaged Navier-Stokes momentum equations for an incompressible, Newtonian fluid are given by~\citep{Reynolds1895}

\begin{equation}\label{eq:RANS_tau}
    U_i \frac{\partial U_j}{\partial x_i} = -\frac{1}{\rho} \frac{\partial P}{\partial x_j} + \nu \frac{\partial^2 U_j}{\partial x_i \partial x_i} -\frac{\partial \tau_{ij}}{\partial x_i} \ , 
\end{equation}
where $U_j$ is the mean velocity, $P$ is the mean pressure, $\rho$ is the fluid density, $\nu$ is the kinematic viscosity, and $\tau_{ij}$ is the Reynolds stress tensor. 

The continuity equation also applies to this flow:

\begin{equation}\label{eq:continuity}
    \frac{\partial U_i}{\partial x_i} = 0 \ .
\end{equation}
Together, Equations~\ref{eq:RANS_tau} and~\ref{eq:continuity} are unclosed. In the most general case, there are four equations (continuity + 3 momentum), and 10 unknowns: $P$, 3 components of $U_i$, and 6 components of $\tau_{ij}$. The goal of turbulence closure modelling is to express $\tau_{ij}$ in terms of $P$ and $U_i$.

The Reynolds stress tensor can be decomposed into isotropic (hydrostatic) and anisotropic (deviatoric) components:

\begin{equation}\label{eq:tau_iso_aniso}
    \tau_{ij} = \frac{2}{3}k\delta_{ij} + a_{ij} \ ,
\end{equation}
where $k = \frac{1}{2}\tau_{kk}$ is half the trace of the Reynolds stress tensor, and $a_{ij}$ is the anisotropy tensor. The isotropic component ($\frac{2}{3}k\delta_{ij}$) in Equation~\ref{eq:tau_iso_aniso} can be absorbed into the pressure gradient term in Equation~\ref{eq:RANS_tau} to form a modified pressure:

%RANS Equation with k taken out (in terms of a)

\begin{equation}\label{eq:RANS_aij}
    U_i \frac{\partial U_j}{\partial x_i} = -\frac{1}{\rho} \frac{\partial}{\partial x_j} \left(P+\frac{2}{3}\rho k\right) + \nu \frac{\partial^2 U_j}{\partial x_i \partial x_i} -\frac{\partial a_{ij}}{\partial x_i} \ . 
\end{equation}

Equation~\ref{eq:RANS_aij} is the form of the RANS momentum equation commonly used in RANS turbulence closure modelling. Several data-driven closure modelling frameworks are based on modelling the Reynolds stress tensor itself, or it's divergence (as in \citet{Brener2022}), which imply the use of Equation~\ref{eq:RANS_tau}. In our work, we model the Reynolds stress anisotropy tensor $a_{ij}$, implying the use of Equation~\ref{eq:RANS_aij}. In either case, the term ``injection" refers to using a model prediction as the closure term in the momentum equation and numerically solving the momentum equation with this predicted closure term.

In an eddy viscosity hypothesis, the anisotropy tensor $a_{ij}$ is postulated to be a function of the mean strain rate $S_{ij}$ and rotation rate $R_{ij}$ tensors:

\begin{align}
 a_{ij} &= a_{ij}(S_{ij}, R_{ij}) \ ,\label{eq:evh}\\
    S_{ij} &= \frac{1}{2}\left(\frac{\partial U_i}{\partial x_j} + \frac{\partial U_j}{\partial x_i}\right) \ , \\
    R_{ij} &= \frac{1}{2}\left(\frac{\partial U_i}{\partial x_j} - \frac{\partial U_j}{\partial x_i}\right) \ .
\end{align}

The eddy viscosity hypothesis  implies the following important constraints:
\begin{itemize}
    \item The anisotropy tensor can be predicted locally and instantaneously. In Equation~\ref{eq:evh}, there is no temporal dependence.
    \item The anisotropic turbulent stresses are caused entirely by mean velocity gradients. 
\end{itemize}
There are many examples of turbulent flows where these assumptions are violated~\citep{Pope2000}. For example, history effects occur in boundary layers~\citep{Bobke2017}. Nevertheless these approximations remain widely used in eddy viscosity modelling. The most common eddy viscosity hypothesis is the linear eddy viscosity hypothesis: 

\begin{equation}
    a_{ij} = - 2 \nu_t S_{ij} \ ,
\end{equation}
where $\nu_t$ is the eddy viscosity, a scalar. This linear hypothesis draws direct analogy from the stress-strain rate relation for a Newtonian fluid. Along with the important constraints implied by invoking an eddy viscosity hypothesis, the linear eddy viscosity hypothesis implies the following:

\begin{itemize}
    \item The anisotropy tensor is aligned with the mean strain rate tensor.
    \item The mapping between mean strain rate and anisotropic stress is isotropic, in that it can be represented using a single scalar ($\nu_t$). 
\end{itemize}

More general non-linear eddy viscosity hypotheses have been used in several models. The primary advantage of these models is that they permit a misalignment of the principal axes of $a_{ij}$ and $S_{ij}$, which occurs for even simple flows. Applying Cayley-Hamilton theorem \citep{,Hamilton1853,cayley1858} to Equation~\ref{eq:evh}, \citet{Pope1975} derived the most general expression for a non-linear eddy viscosity model:

\begin{equation}
    a_{ij} = 2k \sum_{n=1}^{10} g_n \hat{T}^{(n)}_{ij} \ ,
\end{equation}
where $n=1,2,...10$ indexes the scalar coefficients $g_n$, and the following basis tensors:

\begin{align*}
	&\hat{T}^{(1)}_{ij} =\hat{S}_{ij}\ , && \hat{T}^{(6)}_{ij}= \hat{R}_{ik} \hat{R}_{kl} \hat{S}_{lj}+ \hat{S}_{ik} \hat{R}_{kl} \hat{R}_{lj} - \tfrac{2}{3} \hat{S}_{kl}\hat{R}_{lm}\hat{R}_{mk} \delta_{ij}   \ ,\\
	&\hat{T}^{(2)}_{ij} =\hat{S}_{ik} \hat{R}_{kj} - \hat{R}_{ik}\hat{S}_{kj}\ , && \hat{T}^{(7)}_{ij} = \hat{R}_{ik} \hat{S}_{kl} \hat{R}_{lm}\hat{R}_{mj} - \hat{R}_{ik}\hat{R}_{kl} \hat{S}_{lm} \hat{R}_{mj}\ ,\\
	&\hat{T}^{(3)}_{ij} =\hat{S}_{ik}\hat{S}_{kj} - \tfrac{1}{3}\hat{S}_{kl}\hat{S}_{lk} \delta_{ij}\ , && \hat{T}^{(8)}_{ij} = \hat{S}_{ik}\hat{R}_{kl}\hat{S}_{lm}\hat{S}_{mj} - \hat{S}_{ik}\hat{S}_{kl} \hat{R}_{lm} \hat{S}_{mj}\ ,  \\
	&\hat{T}^{(4)}_{ij} =\hat{R}_{ik}\hat{R}_{kj} - \tfrac{1}{3}\hat{R}_{kl}\hat{R}_{lk}\delta_{ij}\ ,  && \hat{T}^{(9)}_{ij} =  \hat{R}_{ik} \hat{R}_{kl}\hat{S}_{lm}\hat{S}_{mj}  +   \hat{S}_{ik} \hat{S}_{kl}\hat{R}_{lm}\hat{R}_{mj} - \tfrac{2}{3} \hat{S}_{kl}\hat{S}_{lm} \hat{R}_{mo}\hat{R}_{ok}  \delta_{ij}   \ ,\\
	&\hat{T}^{(5)}_{ij} = \hat{R}_{ik} \hat{S}_{kl} \hat{S}_{lj} - \hat{S}_{ik} \hat{S}_{kl}\hat{R}_{lj}\ , && \hat{T}^{(10)}_{ij} = \hat{R}_{ik}\hat{S}_{kl}\hat{S}_{lm} \hat{R}_{mo} \hat{R}_{oj}  - \hat{R}_{ik}\hat{R}_{kl}\hat{S}_{lm} \hat{S}_{mo} \hat{R}_{oj}\ .
\end{align*}

The tensors $\hat{S}_{ij}$ and $\hat{R}_{ij}$ above are the non-dimensionalized strain rate and rotation rate tensors. In Pope's original work, these tensors are given by 

\begin{align}
    \hat{S}_{ij} &= \frac{k}{\varepsilon} S_{ij} \ , \\ 
    \hat{R}_{ij} &= \frac{k}{\varepsilon} R_{ij} \ . 
\end{align}

However, other normalization constants are possible, and in several non-linear eddy viscosity models, different basis tensors use different normalization constants. To our knowledge, varying the basis tensor normalization has not been explored in data-driven turbulence closure modelling.

With Pope's general expression for the anisotropy tensor, Equation~\ref{eq:RANS_aij} becomes 

%RANS Equation with b (original TBNN, Kaandorp, etc.)

\begin{equation}\label{eq:RANS_explicit_bij}
    U_i \frac{\partial U_j}{\partial x_i} = -\frac{1}{\rho} \frac{\partial}{\partial x_j} \left(P+\frac{2}{3}\rho k\right) + \nu \frac{\partial^2 U_j}{\partial x_i \partial x_i} -\frac{\partial }{\partial x_i}\left(2k  \sum_{n=1}^{10} g_n \hat{T}^{(n)}_{ij}\right) \ . 
\end{equation}

This is the form of the momentum equation used in several studies which aim to augment the closure relationship via machine learning. For example, in Ling et al.'s ``tensor basis neural network" (TBNN) investigation~\citep{Ling2016}, Equation~\ref{eq:RANS_explicit_bij} was used. \citet{Kaandorp2018}, and \citet{Kaandorp2020} also used this form of the momentum equation. TBNN-based approaches are advantageous for several reasons. The TBNN architecture is motivated by a tensor algebra-based argument that if the anisotropy tensor is to be expanded in terms of the mean strain and rotation rate tensors (a dominant approach in eddy viscosity modelling), the TBNN is the most general form of this expansion~\citep{Pope1975}. Additionally, other techniques to incorporate equivariance are cumbersome, requiring learning and predicting in an invariant eigenframe of some tensorial quantity~\citep{Wu2018,Brener2022}. TBNN based architectures do not require computing eigenvalues of the strain rate tensor at every evaluation data point. Lastly, this closure term is highly expressive, in that ten different combinations of $S_{ij}$ and $R_{ij}$ can be used to represent the anisotropy tensor. However, a major disadvantage with numerically solving Equation~\ref{eq:RANS_explicit_bij} is that the closure term is entirely explicit, greatly reducing numerical stability and conditioning~\citep{Brener2021}. For this reason, \citet{Kaandorp2020} needed to implement a blending function, which blends the fully explicit closure term in Equation~\ref{eq:RANS_explicit_bij} with the more stable implicit closure term treatment made possible with a linear eddy viscosity hypothesis. Here, implicit treatment refers to the way the velocity discretization matrix is constructed in the numerical solver. When a term is treated implicitly, it contributes to the $U_i$ coefficient matrix. In contrast, explicit treatment refers to treating a term as a fixed source term in the discretized equation. Assuming the closure term takes the form $a_{ij} = -2 \nu_t S_{ij}$, Equation~\ref{eq:RANS_aij} can be written as

%Typical RANS Equation (implicit treatment)
\begin{equation}\label{eq:RANS_levm_imp}
    U_i \frac{\partial U_j}{\partial x_i} = -\frac{1}{\rho} \frac{\partial}{\partial x_j} \left(P+\frac{2}{3}\rho k\right) + \frac{\partial}{\partial x_i} \left[(\nu + \nu_t) \frac{\partial U_j}{\partial x_i}\right]  . 
\end{equation}

Equation~\ref{eq:RANS_levm_imp} has the major advantage of increasing diagonal dominance of the $U_j$ coefficient matrix obtained from discretization of this equation, via the eddy viscosity. However, this closure framework only permits the inaccurate linear eddy viscosity closure approximation.

In the present work, we propose the following hybrid treatment of the closure term:
%PoF RANS Equation (implicit treatment + non-linear term)
%Proposed RANS Equation

\begin{equation}\label{eq:RANS_new}
    U_i \frac{\partial U_j}{\partial x_i} = -\frac{1}{\rho} \frac{\partial}{\partial x_j} \left(P+\frac{2}{3}\rho k\right) + \frac{\partial}{\partial x_i} \left[(\nu + \nu_t) \frac{\partial U_j}{\partial x_i}\right] -\frac{\partial }{\partial x_i}\left(2k \sum_{n=2}^{10} g_n \hat{T}^{(n)}_{ij}\right) \ . 
\end{equation}
In Equation~\ref{eq:RANS_new}, the $n=1$ (linear) term has been separated and receives implicit treatment, while the remaining $n = 2,3,...10$ terms grant an opportunity for a machine learning model to provide rich representation of the non-linear part of $a_{ij}$. As we will discuss, the separation of the linear term requires special treatment during training and closure term injection.

\begin{figure}
    \centering
    \includegraphics{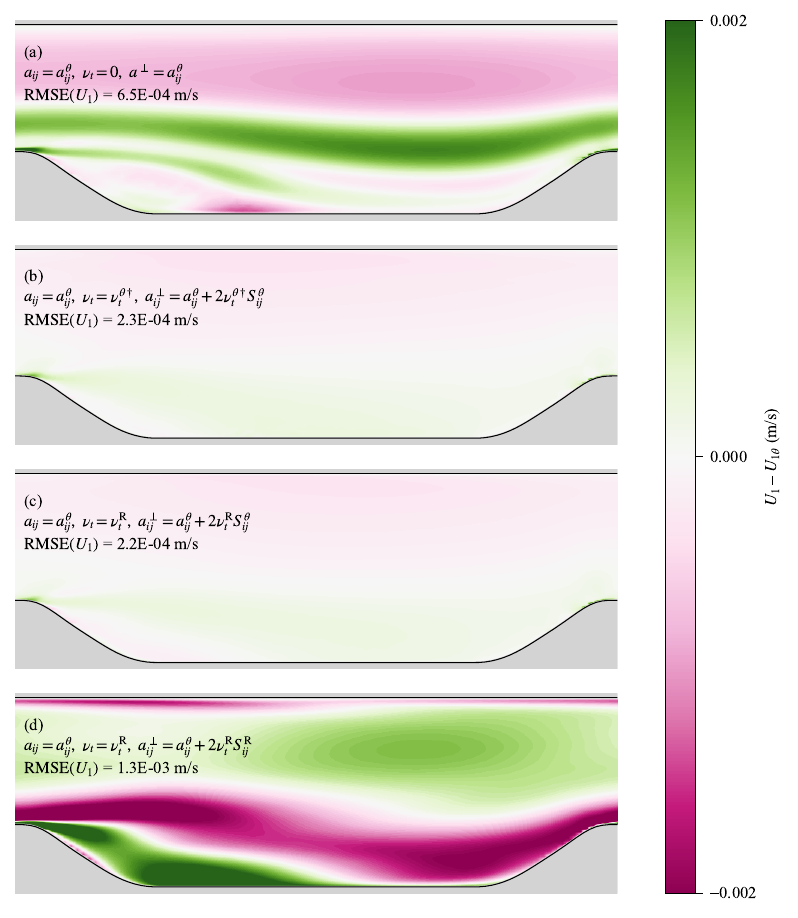}
    \caption[Conditioning test comparing the errors in $U_1$ after injecting various decompositions of the DNS anisotropy tensor into the RANS equations.]{Conditioning test comparing the errors in $U_1$ after injecting: (a) the DNS anisotropy tensor fully explicitly, (b) the optimal eddy viscosity (implicitly), and the remaining non-linear part of the anisotropy tensor calculated using $S^\theta_{ij}$ (explicitly), (c) the eddy viscosity estimated by the $k$-$\omega$ SST model (implicitly), and the remaining non-linear part of the anisotropy tensor calculated using $S^\theta_{ij}$ (explicitly), and (d) the eddy viscosity estimated by the $k$-$\omega$ SST model (implicitly), and the remaining non-linear part of the anisotropy tensor calculated using $S^{\rm R}_{ij}$ (explicitly).}
    \label{fig:conditioning_U}
\end{figure}

\begin{figure}
    \centering
    \includegraphics{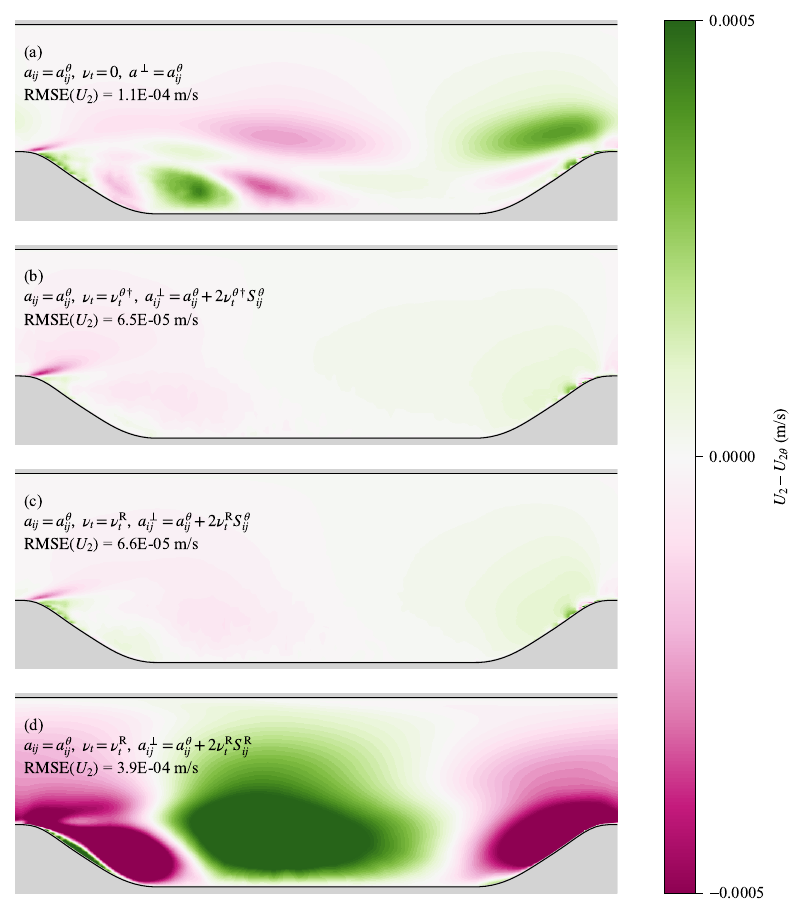}
    \caption[Conditioning test comparing the errors in $U_2$ after injecting various decompositions of the DNS anisotropy tensor into the RANS equations.]{Conditioning test comparing the errors in $U_2$ after injecting: (a) the DNS anisotropy tensor fully explicitly, (b) the optimal eddy viscosity (implicitly), and the remaining non-linear part of the anisotropy tensor calculated using $S^\theta_{ij}$ (explicitly), (c) the eddy viscosity estimated by the $k$-$\omega$ SST model (implicitly), and the remaining non-linear part of the anisotropy tensor calculated using $S^\theta_{ij}$ (explicitly), and (d) the eddy viscosity estimated by the $k$-$\omega$ SST model (implicitly), and the remaining non-linear part of the anisotropy tensor calculated using $S^{\rm R}_{ij}$ (explicitly).}
    \label{fig:conditioning_V}
\end{figure}

\subsection{Conditioning analysis}\label{sec:conditioning_ri}
Various decompositions of the Reynolds stress tensor were investigated by Brener et al.'s conditioning analysis~\citep{Brener2021}. Conditioning analysis for data-driven turbulence closure frameworks is important, since ill-conditioned momentum equations have the potential to amplify errors in the predicted closure term. \citet{Brener2021} concluded that an optimal eddy viscosity approach is necessary to achieve a well-conditioned solution, since it incorporates information about the DNS mean velocity field. In the present work, we demonstrate that the following closure decomposition:

\begin{equation}\label{eq:proposed_decomposition}
    a_{ij} = -2\nu^\text{R}_t S^\theta_{ij} + a^\perp_{ij}
\end{equation}
also achieves a well-conditioned solution. We follow the nomenclature of \citet{Duraisamy2019} in that the $\theta$ superscript indicates a quantity that comes from a high-fidelity source such as DNS. The $\text{R}$ superscript indicates a quantity taken from the corresponding baseline RANS simulation. 

The decomposition in Equation~\ref{eq:proposed_decomposition} permits an augmented turbulence closure framework that treats the machine learning correction only in an explicit term in the momentum equation. Separating the machine learning model prediction has several advantages. It allows the model correction to be easily ``turned off" in unstable situations. It also is more interpretable---rather than correcting both the eddy viscosity and also injecting an explicit correction term, the correction is contained entirely within an explicit term in the momentum equation. Lastly, it avoids the necessity for an optimal eddy viscosity to be computed from high-fidelity data. Though there are methods to increase the practicality of computing the optimal eddy viscosity~\citep{McConkey2022}, this quantity is often unstable and difficult to predict via a machine learning model.

Figures~\ref{fig:conditioning_U} and~\ref{fig:conditioning_V} demonstrate a conditioning test similar to the tests conducted by \citet{Brener2021}. Several different decompositions of the DNS Reynolds stress tensor are injected into a RANS simulation, to identify which decompositions permit a well-conditioned solution. Comparing subfigures (b) and (c) in both Figures~\ref{fig:conditioning_U} and~\ref{fig:conditioning_V}, we can see that the proposed decomposition of the Reynolds stress tensor (Equation~\ref{eq:proposed_decomposition}) achieves an equally well-conditioned solution as the optimal eddy viscosity framework. To our knowledge, this is the first result in the literature showing that a well-conditioned solution can be achieved without the use of an optimal eddy viscosity to incorporate information about the DNS velocity field~\citep{Brener2021}. We further confirm the findings of \citet{Brener2021} with respect to the requirement that the closure decomposition includes information about the DNS mean velocity field in order to achieve a well-conditioned solution. We confirm the notion from 
\citet{Wu2019a} that implicit treatment helps address the ill-conditioning issue, but using the DNS strain rate tensor $S^\theta_{ij}$ to calculate the explicitly injected $a^\perp_{ij}$.
%Conditioning test with proposed decomposition

\subsection{Realizability-informed training}\label{sec:realizability_informed_training}
The Reynolds stress tensor is symmetric positive semidefinite. A set of constraints on the non-dimensional anisotropy tensor $b_{ij}$ arise from this property, as determined by \citet{Banerjee2007}. These constraints are:

%Realizability criteria (Banerjee)
\begin{align}
    -\frac{1}{3} \leq b_{ij} &\leq \frac{2}{3} \ , i= j \ ,\\
    -\frac{1}{2} \leq b_{ij} &\leq \frac{1}{2} \ , i\neq j \ ,\\
    \lambda_1 &\geq \frac{3|\lambda_2| - \lambda_2}{2} \ ,\\
    \lambda_1 &\leq \frac{1}{3}- \lambda_2 \ ,
\end{align}

where the nondimensional anisotropy tensor $b_{ij}$ is calculated by
\begin{equation}
    b_{ij} = \frac{a_{ij}}{2k} \ , 
\end{equation}
and the eigenvalues of $b_{ij}$ are given by $\lambda_1 \geq \lambda_2 \geq \lambda_3$.

A given Reynolds stress tensor is physically realizable if it satisfies these constraints. While the physical realizability of the closure term may seem an important constraint for turbulence models, many commonly used turbulence models such as the  $k$-$\varepsilon$ model~\citep{Launder1974}, $k$-$\omega$~\citep{Wilcox1988}, and $k$-$\omega$ shear stress transport (SST) model~\citep{Menter1994,Menter2003} do not guarantee physical realizability.

Based on the widespread acceptance and popularity of non-realizable turbulence models, it is fair to say that realizability is not a hard constraint on a new turbulence model. Nevertheless, the true Reynolds stress tensor is realizable, and if a machine learning model is able to learn to predict realizable closure terms, it may be more physically accurate. Unfortunately, Pope's tensor basis expansion for the anisotropy tensor (and machine learning architectures based on this expansion) do not provide a means to achieve realizability. To enforce realizability, a variety of ad-hoc strategies have been used, including in the original TBNN paper by \citet{Ling2016}. Most of these strategies involve postprocessing predictions by the TBNN, such as shrinking the predicted anisotropy tensor in certain directions until it is physically realizable~\citep{Jiang2021}.

In the present work, we propose including a penalty for violating realizability constraints in the loss function. In a similar spirit of physics-informed neural networks (PINNs)~\citep{RaissiPINN}, we term the use of this loss function ``realizability-informed training". Whereas PINNs encourage the model to learn physics by penalizing violations of conservation laws, realizability-informed TBNN's learn to predict physically realizable anisotropy tensors.

%Realizability penalty (equations)
The realizability penalty $\mathcal{R}(b_{ij})$ is given as follows:

\begin{equation}\label{eq:realiz_penalty}
\begin{split}
    \mathcal{R}(\tilde b_{ij}) = \frac{1}{6}\left \lbrace  \sum_{i=j} \left( \text{max}\left[\tilde b_{ij}-\frac{2}{3}, - \left(\tilde b_{ij}+\frac{1}{3}\right), 0\right] \right)^2 \right. \\+ \left. \sum_{i\neq j} \left(\text{max}\left[\tilde b_{ij}-\frac{1}{2}, - \left(\tilde b_{ij}+\frac{1}{2}\right), 0\right]\right)^2  \right \rbrace \\+ \frac{1}{2}\left \lbrace \left(\text{max}\left[\frac{3|\lambda_2| - |\lambda_2|}{2} - \lambda_1, 0 \right]\right)^2 \right.+ \\\left. \left(\text{max}\left[\lambda_1 - \left( \frac{1}{3}-\lambda_2\right),0 \right] \right)^2  \right \rbrace \ ,
\end{split}
\end{equation}
where the $\sim$ above a symbol denotes a model prediction for the quantity associated with the symbol.

\begin{figure}
\centering
\begin{subfigure}[b]{\textwidth}
    \centering
    \includegraphics{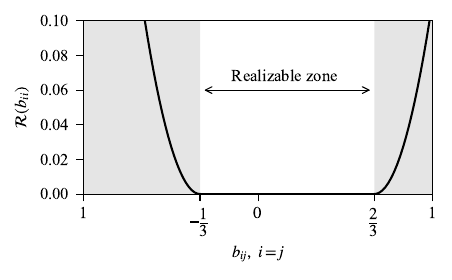}
    \caption{}
\end{subfigure}
\hfill
\begin{subfigure}[b]{\textwidth}
    \centering
    \includegraphics{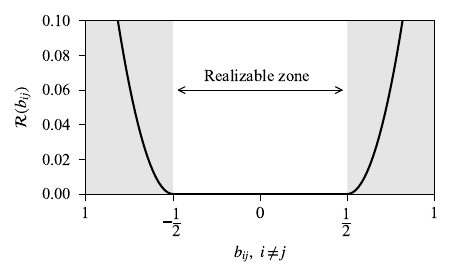}
    \caption{}
\end{subfigure}
\hfill
\caption[Realizability penalty for the diagonal and  off-diagonal components of $b_{ij}$.]{Realizability penalty for the (a)  diagonal components of $b_{ij}$, and (b)  off-diagonal components of $b_{ij}$.}\label{fig:realiz_penalty}
\end{figure}

Equation~\ref{eq:realiz_penalty} can be thought of as the mean squared violation in the components of $\tilde b_{ij}$, plus the mean squared violation in the eigenvalues of $\tilde b_{ij}$. To help visualize this penalty function, figure~\ref{fig:realiz_penalty} shows the penalties incurred by violating various realizability constraints on the anisotropy tensor.

%Realizability penalty (plots)

It should be noted that in the same way that a PINN's prediction cannot be guaranteed to satisfy a conservation law, a realizability-informed TBNN cannot be guaranteed to predict a physically realizable anisotropy tensor. The goal is that the incorporation of a physics-based loss at training time will encode into the model a tendency to predict physically realizable anisotropy tensors. At training time, when the model predicts a $b_{ij}$ component outside the realizability zone, then it is penalized in two ways: the error in this prediction will be non-zero (since all label data are realizable), and the realizability penalty will be non-zero. Therefore, the realizability penalty serves as an additional driving factor towards the realizable, true $b_{ij}$ value. The merits of this approach are demonstrated in Section~\ref{sec:realizability_results}.

%Modified error in loss function - plot of b vs T vs a
Another important metric in the loss function is the error-based penalty $\mathcal{E}(\tilde b_{ij})$. When the model predicts a certain anisotropy tensor, it is evaluated against a known high-fidelity anisotropy tensor available in the training dataset. Typically, mean-squared error loss functions are used to train machine learning augmented closure models. However, we propose the following modifications to the loss function of a TBNN:
%Modified error in loss function - equation
\begin{enumerate}
\item Since $b_{ij}$ is a symmetric tensor, we propose to sum the squared errors as follows:
    \begin{equation}\label{eq:mse_b}
        \mathcal{E}(\tilde b_{ij}) = \frac{1}{6}\left\lbrace \sum_{\substack{ ij \in \lbrace 11, 12, 13,\\ 22,23,33\rbrace}} (\tilde b_{ij} - b^\theta_{ij})^2 \right \rbrace \ .
    \end{equation}
    Calculating the squared-error in this way avoids double-penalizing the off-diagonal components, a situation which arises when summing over all components of $b_{ij}$.
    \item Though the TBNN model predicts $b_{ij}$, we propose to use a loss function based on the error in $a_{ij}$. Near the wall, $\hat{T}^{(n)}_{ij}\rightarrow0$, but $b_{ij} \nrightarrow 0$.

    The vanishing of $\hat{T}^{(n)}_{ij}$ with non-vanishing $b_{ij}$ causes instabilities during training, leading to $g_n \rightarrow \infty$ here. Since $a_{ij}$ is the tensor injected into the momentum equation, its accurate prediction should be the focus of the training process. To dimensionalize $b_{ij}$, the turbulent kinetic energy $k$ must be used. This leads to the following error-based loss:
    \begin{equation}
        \mathcal{E}(\tilde a_{ij}) = \frac{1}{6}\left\lbrace \sum_{\substack{ ij \in \lbrace 11, 12, 13,\\ 22,23,33\rbrace}} ( 2k^\theta \tilde b_{ij} - 2k^\theta b^\theta_{ij})^2 \right  \rbrace =  \frac{2(k^\theta)^2}{3}\left\lbrace \sum_{\substack{ ij \in \lbrace 11, 12, 13,\\ 22,23,33\rbrace}} (\tilde  b_{ij} - b^\theta_{ij})^2 \right  \rbrace \ , \label{eq:mse_a_long}
    \end{equation}
    or stated more simply, 
    \begin{equation}\label{eq:mse_a_short}
         \mathcal{E}(\tilde a_{ij})=(2k^\theta)^2 \mathcal{E}(\tilde b_{ij}) \ .
    \end{equation}
    The reason for the use of $k^\theta$ in Equation~\ref{eq:mse_a_long} and Equation~\ref{eq:mse_a_short} is discussed in Section~\ref{sec:NN_architecture}.
    
\end{enumerate}

The final loss function includes an error-based metric and a realizability-violation penalty. This loss function $\mathcal{L}$ is given by

%Final loss function
%\begin{equation}\label{eq:loss}
%    \mathcal{L} = \frac{1}{N}\sum_{p} \frac{(2k^\theta_p)^2}{\left(\overline{||a_{ij,{p\in \text{case}(p)}}||}\right)^2}\left( \mathcal{E}(\tilde b_{ij,p})+\alpha \mathcal R(\tilde b_{ij,p})\right) \ ,
%\end{equation}
\begin{equation}\label{eq:loss}
    \mathcal{L} = \frac{1}{N}\sum_{p} \frac{(2k^\theta_p)^2}{\displaystyle \sum_{k=1}^s Z_k^2 {\mathcal I}_{C_k}(p)}\left( \mathcal{E}(\tilde b_{ij,p})+\alpha \mathcal R(\tilde b_{ij,p})\right) \ ,
\end{equation}
where $\alpha$ is a factor used to control the relative importance of the realizability penalty. $N$ is the total number of points in the dataset. The points are indexed by $p = 1,2,\ldots,N$. The cases in the dataset are indexed $k=1,2,\ldots,s $. ${\mathcal I}_{C_k}(p)$ is an indicator function:
\begin{equation}
    {\mathcal I}_{C_k}(p) = \begin{cases}
1, \text{if } p \in C_k,\\
0 \ \text{otherwise.}
\end{cases}
\end{equation}

For a given point $p$, ${\mathcal I}_{C_k}(p)$ selects all points which come from the same case as $p$. The denominator for all points from the same case $C_k$ is the same---normalization is applied on a case-by-case basis. A given point is normalized by the mean-squared Frobenius norm of the anisotropy tensor over all the points from the case it comes from. The mean Frobenius norm over a case is given by:

$$Z_k = \frac {1}{|C_k|}\sum_{p \in C_k} \| a_{ij}(p)\|_F\ ,\qquad k=1,2,\ldots,s\ ,$$
where $|C_k|$ is the cardinality of the case $C_k$ (viz., the number of points in the $k$th case $C_k$). $Z_k$ is simply the average of the Frobenius norm $\|\ \ \cdot\ \ \|_F$ of the anisotropy tensor $a_{ij}$ for all points $p$ (viz., $a_{ij}(p)$) in case $C_k$. The objective of this denominator in Equation~\ref{eq:loss} is to promote a more balanced regression problem, since data points from various cases or flow types may have $||a_{ij}||$ that differ by orders of magnitude. Normalization on a case-by-case basis is made in an effort to normalize all $a_{ij}$ error magnitudes to a similar scale.

In this study, we use $\alpha=10^2$ to encode a high preference for realizable results in Section~\ref{sec:results_ri}. Lower values of $\alpha$ will reduce the penalty applied to realizability violations, which may be necessary for flows in which the anisotropy tensor is difficult to predict via a TBNN. As discussed, the multiplicative term $(2k^\theta)^2$ is used to formulate the loss function in terms of predicting $a_{ij}$ rather than $b_{ij}$. However, $\mathcal{R}(\tilde b_{ij})$ is also multiplied by $(2k^\theta)^2$ to ensure that the realizability penalty and mean-squared error in $a_{ij}$ are of similar scales.

\subsection{Neural network architecture}\label{sec:NN_architecture}
Motivated by improving the training and injection stability, as well as generalizability, we propose several modifications to the original tensor basis neural network (TBNN) \citep{Ling2016}. 

\begin{figure}
    \centering
    \includegraphics[width=\textwidth]{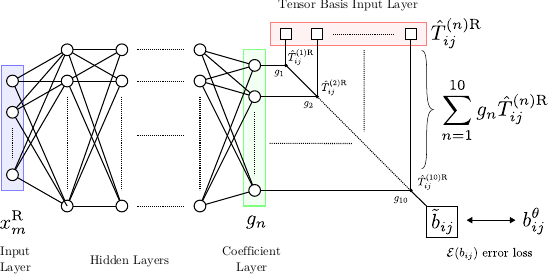}
    \caption{Model architecture and training configuration for the original TBNN proposed by \citet{Ling2016}.}
    \label{fig:tbnn_vanilla}
\end{figure}

The original TBNN is shown in figure~\ref{fig:tbnn_vanilla}. At training time, this network predicts the non-dimensional anisotropy tensor $b_{ij}$. All basis tensors used in this prediction at training time come from RANS, and during training the prediction $\tilde b_{ij}$ is evaluated against a known value of $b^\theta_{ij}$ from a high-fidelity simulation. At injection time, the network is used in this same configuration to predict $\tilde b_{ij}$. $\tilde b_{ij}$ is injected into a coupled system of equations consisting of the continuity/momentum equations (explicit injection), as well as the turbulence transport equations. This system of equations is iterated around a fixed $\tilde b_{ij}$, to obtain an updated estimate for the turbulent kinetic energy $k$, and therefore an updated estimate for $\tilde a_{ij} = 2 k \tilde b_{ij}$. 

\begin{figure}
    \centering
    \includegraphics[width=\textwidth]{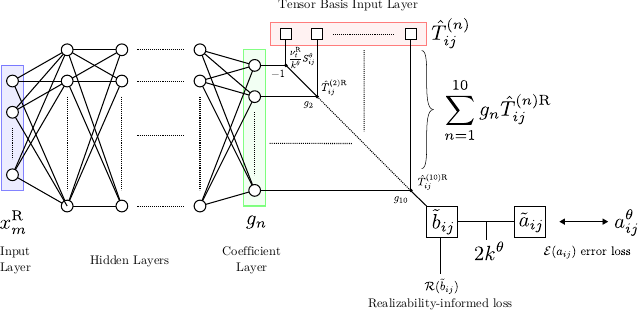}
    \caption[Model architecture and training configuration proposed in the present investigation.]{Model architecture and training configuration proposed in the present investigation. Note that during training, $S^\theta_{ij}$ is used in $\hat{T}^{(1)}_{ij}$, whereas all other quantities come from the original RANS simulation.}
    \label{fig:tbnn_training}
\end{figure}

The modified TBNN is shown in figure~\ref{fig:tbnn_training}. This TBNN relies on the same tensor basis expansion as the original TBNN. However, the linear term has been modified in this expansion. Whereas the original TBNN uses

\begin{equation}
    \hat{T}^{(1)}_{ij} = \frac{k^\text{R}}{\varepsilon^\text{R}}S^\text{R}_{ij} \ , 
\end{equation}
our modified TBNN uses

\begin{equation}
    \hat{T}^{(1)}_{ij} = \frac{\nu_t^\text{R}}{k^\theta}S^\theta_{ij} \ . 
\end{equation}

Further, while the original TBNN calculates the linear (denoted by superscript $\text{L}$) component of $b_{ij}$ as 

\begin{equation}
    b^{\text{L}}_{ij} = g_1 \hat{T}^{(1)}_{ij} = g_1\frac{k}{\varepsilon}S_{ij} \ ,
\end{equation}
our modified TBNN uses 

\begin{equation}\label{eq:proposed_linear}
    b^{\text{L}}_{ij} = - \hat{T}^{(1)}_{ij} = - \frac{\nu_t}{k}S_{ij} \ ,
\end{equation}
where the superscript $\text{R}$ denotes a quantity that comes from the original RANS simulation.

These changes are motivated by the following:

\begin{enumerate}
    \item At injection time, we use implicit treatment of the linear term $\hat{T}^{(1)}_{ij}$ to formulate Equation~\ref{eq:RANS_new} in a stable manner. After injection, $S_{ij}$ will continue to evolve. In a similar spirit as optimal eddy viscosity frameworks, we therefore use $S^\theta_{ij}$ to compute the linear component at training time. In optimal eddy viscosity based frameworks, using $S_{ij} = S^\theta_{ij}$ at training time helps drive $U_j\rightarrow U^\theta_{j}$ at injection time (the cause of this behavior is currently unknown). In \citet{Ling2016}, all basis tensors (and therefore $b_{ij}$) remain fixed after injection. We also fixed $\nu_t$ = $\nu^\text{R}_t$ at injection time (viz., no further evolution of the eddy viscosity is permitted).
    \item At training time, we dimensionalize $\tilde b_{ij}$ using $k^\theta$: $\tilde a_{ij} = 2k^\theta \tilde b_{ij}$. At evaluation time, we do not have $k^\theta$. However, the need for $k^\theta$ is avoided, since \textit{we only use the non-linear part of $\tilde b_{ij}$ at test time}. The reason we only need $\tilde b^\perp_{ij}$ at test time is that the training process has been designed to use the linear part of $b_{ij}$ estimated by the RANS turbulence model, and augment this by the TBNN's equivariant prediction for $\tilde b^\perp_{ij}$.
    \item Using $\nu_t/k$ to normalize the basis tensors and fixing $g_1=-1 $ in Equation~\ref{eq:proposed_linear} results in the RANS prediction for $b^\text{L}_{ij}$ being implicitly used in the TBNN. Therefore, the TBNN learns to correct $b_{ij}$ using $b^\perp_{ij}$ in a way that allows a realizability-informed training process, and fully implicit treatment of the linear term at injection time.
\end{enumerate} 
%Figure - NN architecture with fixed g1, separate NN for k correction

\begin{figure}
    \centering
    \includegraphics[width=\textwidth]{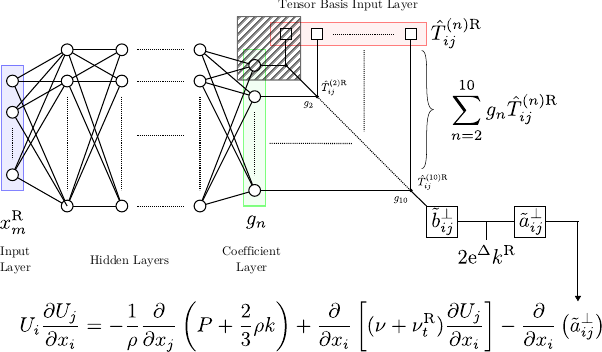}
    \caption[Test-time injection configuration proposed in the present investigation. ]{Test-time injection configuration proposed in the present investigation. The hatching over $g_1$ and $\hat{T}^{(1)}_{ij}$ indicates that they are not used at injection time. }
    \label{fig:tbnn_testing}
\end{figure}

Lastly, the use of $k^\theta$ to dimensionalize $\tilde b_{ij}$ is also enabled by our use of a separate neural network to correct $k^\text{R}$ at injection time. This neural network is called the $k$-correcting neural network (KCNN), and is a simple fully-connected feed-forward neural network that predicts a single output scalar $\Delta$:

\begin{equation}
    \Delta  = \log \left(\frac{k^\theta}{k^\text{R}}\right) \ ,
\end{equation}
such that at injection time, an updated estimate for $k$ can be obtained $\tilde k = {\rm e} ^\Delta k^\text{R}$, without the need to re-couple the turbulence transport equations. The KCNN shares the same input features as the TBNN. 

Together, the KCNN and TBNN predict the anisotropy tensor in the following manner:

\begin{equation}
    \tilde a_{ij} = 2 \left({\rm e}^\Delta k^\text{R} \right) \sum_{n=1}^{10}g_{n} \hat{T}^{(n)}_{ij} \ .
\end{equation}

With the linear component of $\tilde a_{ij}$ being treated implicitly during injection, the entire closure framework is summarized as explicit injection of the following term into the momentum equation:

\begin{equation}
    \tilde a^\perp_{ij} = 2 \left({\rm e}^\Delta k^\text{R} \right) \sum_{n=2}^{10} g_{n} \hat{T}^{(n)\text{R}}_{ij} \ .
\end{equation}

\subsubsection{Tensor basis Kolmogorov-Arnold network (TBKAN)}

\begin{figure}
    \centering
    \includegraphics[width=\textwidth]{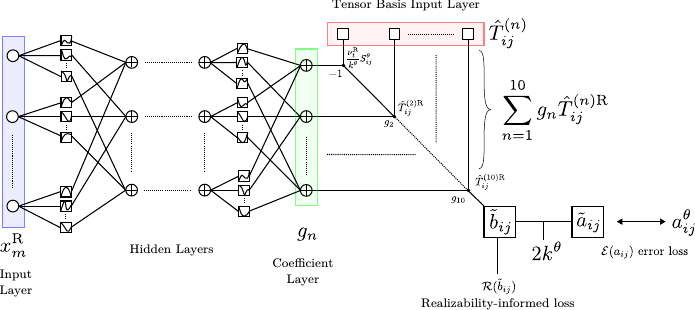}
    \caption{Model architecture and training configuration for TBKAN which replaces the multi-layer perceptron in the modified TBNN (see figure~\ref{fig:tbnn_training}) with a Kolmogorov-Arnold network.}
    \label{fig:tbkan-1}
\end{figure}

The tensor basis Kolmogorov-Arnold network shown in figure~\ref{fig:tbkan-1} in the training configuration replaces the multi-layer perceptron in the modified TBNN with a Kolmogorov-Arnold network (KAN) introduced by \citet{liu2024kan}. The Kolmogorov-Arnold representation theorem states that any continuous multivariate function can be expressed as a composition of continuous univariate functions. KANs are based on this theorem, replacing the typical linear weight matrices in neural networks with learnable one-dimensional (1D) functions \citep{liu2024kan}. These functions are parameterized using splines, offering a flexible and computationally efficient approach to represent continuous functions. KANs utilize B-splines, which are piecewise polynomial functions, to model local variations in data. Each spline segment corresponds to a polynomial function, and their piecewise nature allows KANs to approximate the local variations in the data during training. This adaptability enables KANs to capture intricate functional relationships more effectively, merging the advantages of B-splines with the traditional neural network framework, thereby enhancing both accuracy and interpretability.

In the TBKAN architecture, KAN replaces the hidden layers of the standard TBNN, while the anisotropy mapping portion remains unchanged. The TBKAN’s output layer is designed to predict the coefficients of the tensor basis.

%introduce KAN here
% 1 paragraph introducing KAN, highlighting key features of KAN
% 1 paragraph explaining where KAN is used within TBNN architecture (It basically replaces the hidden layers, the anisotropy mapping part is unchanged).
% Figure in same style as other figures in this section showing KAN instead of MLP. You can even probably just draw the squiggles inside the existing circles? I would download the PDF files and edit them using e.g. inkscape. That way, the vector drawing is preserved. 

%either take this out 

%\begin{figure}
%   \centering
%    \includegraphics[width=\textwidth]{figures/KAN-FP-results/kan-image-2.pdf}
%    \caption{Test-time injection configuration proposed in the present investigation. Test-time injection configuration proposed in the present investigation. The hatching over $g_1$ and $\hat{T}^{(1)}_{ij}$ indicates that they are not used at injection time.}
%    \label{fig:tbkan-2}
%\end{figure}

\subsection{Machine learning procedure}\label{sec:machine_learning}
The KCNN, TBNN, and TBKAN architectures proposed in Section~\ref{sec:NN_architecture} were implemented in PyTorch, along with the proposed realizability-informed loss function (Equation~\ref{eq:loss}). These models were trained using an open-source dataset for data-driven turbulence modelling~\citep{McConkeySciDataPaper2021}. The objective of this study is to train and evaluate models trained on various flows, to determine whether the proposed realizability-informed loss function and architecture modifications are significantly beneficial. All code is available on Github~\citep{tbnn_github}.

\subsubsection{Datasets}\label{sec:datasets}
The training flows consist of flow over periodic hills~\citep{Xiao2020}, flow through a square duct~\citep{Pinelli2010}, and flow over a flat plate with zero pressure gradient~\citep{NASAturbmodel}. These flows are selected because they contain several challenging physical phenomena for RANS, including separation, reattachment, and Prandtl's secondary flows. The flat plate case is also included, to demonstrate how machine learning can improve the anisotropy estimates within the boundary layer. For each flow type, both a hold-out validation set, and a hold-out test set are selected. The validation set is used during training to help guide when to stop training in order to prevent overfitting, but the validation set loss is not back-propagated through the network to update weights and biases. While we hold-out an entire case for the test set (the usual procedure in data-driven turbulence modelling), we also generate the validation sets by holding out entire cases at a time.  We recommend this method for generating validation sets in data-driven turbulence modelling---it is analogous to grouped cross-validation, a practice used in machine learning where several data points come from a single observation. Here, we consider each separate flow case as a single observation, each containing many data points. It is therefore prudent to ensure that two data points from the same observation are not used in both the training and test set. 

\begin{table}
\centering
\fontsize{10}{12}\selectfont
\renewcommand{\arraystretch}{1.25}
\caption[Datasets used for training, validation, and testing. ]{Datasets used for training, validation, and testing. Varied parameters are defined in Section~\ref{sec:generalizability_tests}.}
\label{tbl:datasets}
\begin{tabular}{cccc}
\hline
                                                                       & Flat plate                                                          & Square duct                                                                                                   & Periodic hills \\ \hline
\multicolumn{1}{c|}{\begin{tabular}[c]{@{}c@{}}Data\\ points\end{tabular}}      & 1,396                                                               & 147,456                                                                                                       & 73,755         \\ \hline
\multicolumn{1}{c|}{\begin{tabular}[c]{@{}c@{}}Parameter\\ varied\end{tabular}} & $Re_\theta$                                                         & $Re_H$                                                                                                        & $\alpha$       \\ \hline
\multicolumn{1}{c|}{Training set}                                      & \begin{tabular}[c]{@{}c@{}}1000, 2000, 2540,\\ 3270, 3970\end{tabular} & \begin{tabular}[c]{@{}c@{}}1100, 1150, 1250, \\ 1350, 1400, 1500,\\ 1600, 2205, 2400, \\ 2600, 2900, 3500\end{tabular} & 0.5, 1.0, 1.5  \\ \hline
\multicolumn{1}{c|}{Validation set}                                    & 1410, 3030, 4060                                                    & 1300, 1800, 3200                                                                                              & 0.8            \\ \hline
\multicolumn{1}{c|}{Test set}                                          & 3630                                                                & 2000                                                                                                          & 1.2            \\ \hline
\end{tabular}
\end{table}
Table~\ref{tbl:datasets} outlines the three training/validation/test splits considered. The objective in splitting the dataset this way is to determine whether a realizability-informed model can generalize to a new case for a given flow. Machine learning-based anisotropy mappings do not generalize well to entirely new flows~\citep{Man2023,McConkey2022b,Duraisamy2021}. However, they can be used to dramatically enhance the performance of a RANS simulation for a given flow type. In this same spirit, we aim to test how our modifications improve the generalizability of the learned anisotropy mapping to an unseen flow, albeit within the same class of flow.

\subsubsection{Input features}\label{sec:input_features}
The input features here are all derived from the baseline RANS $k$-$\omega$ SST simulation. The input features form the vector $x^\text{R}_m$. The superscript $\text{R}$ has been dropped in this section to avoid crowded notation, but it applies to all quantities discussed in Section~\ref{sec:input_features}.

The input features must be Galilean invariant in order to generate an appropriately constrained anisotropy mapping. Most data-driven anisotropy mapping investigations use a mixture of heuristic scalars and scalars systematically generated from a minimal integrity basis for a set of gradient tensors~\citep{Wu2018}. We emphasize that all scalars must be Galilean invariant---without this criteria, the RANS equations will lose Galilean invariance. Despite the importance of this constraint, several data-driven anisotropy mappings include scalars like the turbulence intensity, which breaks Galilean invariance.

We use a mixture of heuristic scalars and scalars systematically generated from a minimal integrity basis \citep{Wu2018}. We use the following heuristic scalars:
\begin{align}
    q_1 &= \text{min}\left(\frac{\sqrt{k} y_w}{50 \nu},2\right)\ , \label{eq:q2_ri}\\
	q_2 &= \frac{k}{\varepsilon} \sqrt{\sum_{i} \sum_{j} |S_{ij}|^2}\ , \label{eq:q3_ri}\\
    q_3 &= \frac{\sqrt{\displaystyle\sum_{i} \displaystyle\sum_{j} |\tau_{ij}|^2}}{k} \ , \\
    q_4 &= \frac{\sqrt{k}}{0.09\omega y_w} \ ,\\
    q_5 &= \frac{500\nu}{y_w^2 \omega} \ ,\\
    q_6 &= \text{min}\left(\text{max}\left(q_4,q_5\right),\frac{2.0k}{\text{max}\left(\dfrac{y_w^2}{\omega}\dfrac{\partial k}{\partial x_i}\dfrac{\partial \omega}{\partial x_i}\right)}\right) \ ,
\end{align}
with $\varepsilon = 0.09 \omega$, and $y_w$ is the distance to the nearest wall.

These input features correspond to: the wall-distance based Reynolds number ($q_1$), the ratio of turbulent time scale to mean strain timescale ($q_2$), the ratio of total Reynolds stress to $k$ ($q_3$), and different blending scalars used within the $k$-$\omega$ SST model ($q_4$, $q_5$, $q_6$). 
The full list of integrity basis tensors is given in Appendix~\ref{ap:symbolic_results}. We use the first invariant (the trace, $A_{ii}$) of the following tensors: $B^{(1)}_{ij}$, $B^{(3)}_{ij}$, $B^{(4)}_{ij}$, $B^{(6)}_{ij}$, $B^{(7)}_{ij}$, $B^{(16)}_{ij}$, and $B^{(35)}_{ij}$. We use the second invariant, $\frac{1}{2}\left(\left(A_{ii}\right)^2 - A_{ij}A_{ji}\right)$, of the following tensors: $B^{(3)}_{ij}$, $B^{(6)}_{ij}$, $B^{(7)}_{ij}$, $B^{(8)}_{ij}$.
These input features were hand-picked from the full set of 94 invariants listed in Appendix~\ref{ap:symbolic_results}~\citep{Wu2018,McConkey2022}. As discussed in \citet{McConkey2022}, many of the invariants are zero for 2D flows. However, different conditions cause different invariants to be zero. For example, in the present study, there are a different set of zero invariants for flow through over periodic hills, and flow through a square duct. This difference occurs because different components of $\partial () / \partial x_j$ are to be zero. We have performed a systematic investigation using a symbolic math toolbox (sympy~\citep{sympy}) to determine which invariants are zero for the duct case, and general 2D flows. We make the results and code available in Appendix \ref{ap:symbolic_results} and on Github~\citep{integritybasisgithub}, respectively. Input features used in this investigation were selected based on the results in Appendix~\ref{ap:symbolic_results}. Therefore, the input features are not uniformly zero on any of the considered flows.

To ensure all input features are of the same magnitude during training, they are scaled according to the following formula:

\begin{equation}
    x^\text{R}_m = \frac{X^\text{R}_m - \mu_{m}}{\sigma_m} \ ,
\end{equation}
where $x^\text{R}_m$ is the input feature vector for the neural network, $X^\text{R}_m$ is the raw input feature vector from the RANS simulation, $\mu_m$ is a vector containing the mean of each input feature over the entire training dataset, and $\sigma_{m}$ is a vector containing the standard deviation of each input feature over the entire training dataset.

When making predictions on the hold-out validation and test sets, the mean and standard deviation values from the training data are used to avoid data leakage.

\subsubsection{Hyperparameters and training procedure}\label{sec:hyperparameters_ri}
The hyperparameters for each neural network were hand-tuned based on validation set performance. The hidden layers for all TBNNs and KCNNs are fully connected, feedforward layers with Swish activation functions~\citep{Ramachandran2017}. The appropriate hyperparameters vary between flows, since each dataset contains a different number of data points, and the anisotropy mapping being learned is distinct. Table~\ref{tbl:hyperparameters_ri} shows the final hyperparameters used. All training runs used a mini-batch size of 32, except the periodic hills TBNN run, which used a mini-batch size of 128.

\begin{table}
\centering
\caption{Hyperparameters selected for each model.}
\fontsize{10}{12}\selectfont
\renewcommand{\arraystretch}{1.25}
\label{tbl:hyperparameters_ri}
\begin{tabular}{cccccc}
\hline
Model & Dataset        & Learning rate  & Epochs & \begin{tabular}[c]{@{}c@{}}Hidden\\ layers\end{tabular} & \begin{tabular}[c]{@{}c@{}}Neurons\\ per layer\end{tabular} \\ \hline
TBNN  & Flat plate     & $5(10)^{-4}$   & 7959   & 4                                                       & 20                                                          \\
KCNN  & Flat plate     & $2.5(10)^{-4}$ & 2478   & 5                                                       & 30                                                          \\
TBNN  & Square duct    & $5(10)^{-4}$   & 100    & 7                                                       & 30                                                          \\
KCNN  & Square duct    & $5(10)^{-4}$   & 1150   & 7                                                       & 30                                                          \\
TBNN  & Periodic hills & $1(10)^{-5}$   & 19140  & 7                                                       & 30                                                          \\
KCNN  & Periodic hills & $1(10)^{-6}$   & 13138  & 11                                                      & 30                                                          \\ \hline
\end{tabular}
\end{table}

\begin{figure}
\centering

\begin{subfigure}[b]{0.45\textwidth}
    \centering
    \includegraphics{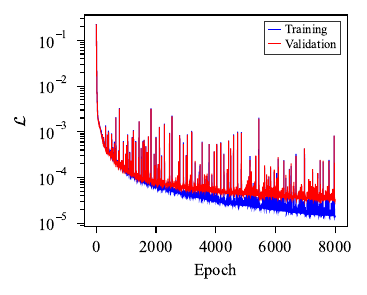}
    \caption{}
\end{subfigure}
\hfill
\begin{subfigure}[b]{0.45\textwidth}
    \centering
    \includegraphics{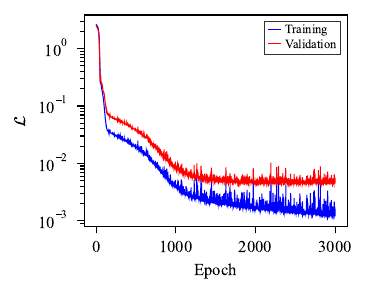}
    \caption{}
\end{subfigure}

\begin{subfigure}[b]{0.45\textwidth}
    \centering
    \includegraphics{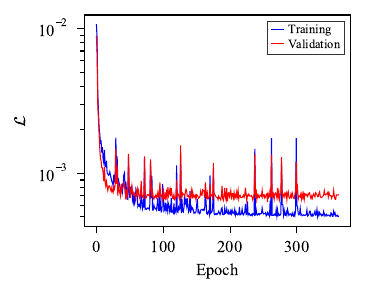}
    \caption{}
\end{subfigure}
\hfill
\begin{subfigure}[b]{0.45\textwidth}
    \centering
    \includegraphics{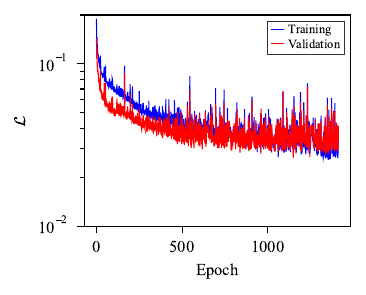}
    \caption{}
\end{subfigure}

\begin{subfigure}[b]{0.45\textwidth}
    \centering
    \includegraphics{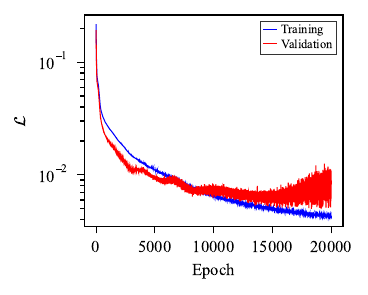}
    \caption{}
\end{subfigure}
\hfill
\begin{subfigure}[b]{0.45\textwidth}
    \centering
    \includegraphics{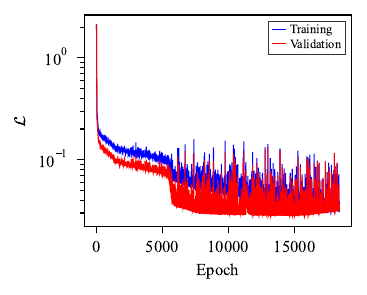}
    \caption{}
\end{subfigure}
\caption[Loss function vs epoch for the TBNN and KCNN models trained on each dataset.]{Loss function vs epoch for the (a) TBNN and (b) KCNN models on the flat plate dataset; the (c) TBNN and (d) KCNN models on the square duct dataset; and the (e) TBNN and (f) KCNN models on the periodic hills dataset.}\label{fig:loss_curves}
\end{figure}

The AMSGrad Adam optimizer~\citep{Reddi2018} was found to achieve better performance than the standard Adam optimizer~\citep{Kingma2015} for training TBNNs. The learning rate and number of epochs for each optimizer is given in table~\ref{tbl:hyperparameters_ri}. Satisfactory performance was achieved with a constant learning rate; for training on more complex flows we recommend the use of learning rate scheduling to achieve better performance. The training/validation loss curves for each model are shown in figure~\ref{fig:loss_curves}.

\subsubsection{Hyperparameter tuning for TBKAN}\label{sec:impact_hyperparameter_tuning}  

The performance of the TBKAN was extensively optimized through a combination of systematic hyperparameter tuning and manual adjustments. The architecture of the TBKAN model was configured as \([6, 9, 10]\), where 6 corresponds to the number of input features, 9 represents the network width (number of neurons per hidden layer), and 10 denotes the output size of the network. The depth of the network, representing the number of hidden layers, was fixed to 2 for the flat plate case.

Hyperparameter tuning focused on refining the grid size (\( g \)), network width (\( w \)), spline order, and input feature combinations. An initial set of 27 runs, conducted with randomly selected configurations, broadly explored the hyperparameter space. These runs were used as a warm start for Bayesian optimization (BO), which further utilized 143 trials to systematically refine the hyperparameters. The configuration employed a grid size of 8 control points for the B-spline basis representation, with the polynomial order $k$ of the splines fixed at a value of three (viz., $k=3$ corresponding to cubic splines). This choice was found to provide the best balance between flexibility and computational efficiency.

\begin{figure}[h]
\centering
\includegraphics{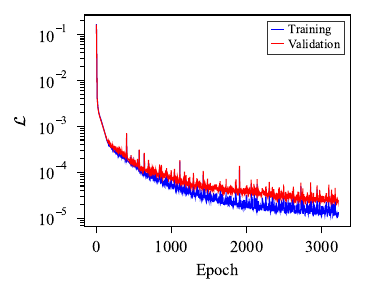}
\caption[Loss function vs epoch for the TBKAN model trained on the flat plate dataset.]{Loss function vs epoch for the TBKAN model trained on the flat plate dataset, with the best-performing KAN configuration given by a network width of $w=9$, a grid size of $g=8$, and a polynomial order of $k=3$ (cubic B-spline).}
\label{fig:loss_curves_KAN}
\end{figure}

The final hyperparameters for the best-performing model were as follows: architecture \([6, 9, 10]\), a depth of 2 layers, a grid size of 8 control points, and a learning rate of \( 4.9 \times 10^{-3} \). This configuration achieved a mean-squared error (MSE) of 0.22 on the flat plate case. Input feature selection was refined by fixing three core features while sampling from a broader set to enhance the model's adaptability in predicting finer details of the anisotropy tensor. The AMSGrad Adam optimizer \citep{Reddi2018} was used for all training runs, with a mini-batch size of 32, which ensured stable convergence. Training and validation loss curves for the best-performing model for the flat plate, corresponding to a KAN configuration with $w=9$, $g=8$, and $k=3$ (cubic splines), are shown in figure~\ref{fig:loss_curves_KAN}.

\subsection{Computational costs} 
The inference-time cost for the proposed methodology varies significantly from case-to-case. A prediction requires 1. running a baseline RANS simulation, 2. evaluating the ML model predictions, and 3. running a corrected RANS simulation. Compared to steps 1 and 3, the cost of step 2 (model inference) is negligible. Additionally, since the converged fields from step 1 are used to initialize the simulation in step 3, the cost of the corrected RANS simulation is also reduced. Generally, we found that combined inference time costs for steps 1, 2, and 3 were between 1.5 and 3 times the cost of the baseline simulation (i.e., steps 1, 2, and 3 cost about 1.5--5 times as much as step 1), depending on how much the injected closure fields differ from the original linear eddy viscosity-based field. The training cost also varies depending on the dataset. For the periodic hills and square duct datasets here, the training time was approximately 16 GPU hours on a single NVIDIA RTX 3090 GPU.

\section{Results for Proposed TBNN and TBKAN architectures}\label{sec:results_ri}

\subsection{Generalization tests}\label{sec:generalizability_tests}
It was of interest to determine how well the trained models generalize to unseen variations of their training flows. This section demonstrates generalization results for flow over a flat plate with zero pressure gradient, flow through a square duct, and flow over periodic hills. 

For all cases, the original RANS solution was generated using OpenFOAM v2212, assuming an isothermal, incompressible, and Newtonian fluid. Simulation parameters such as solver, schemes, and solution methodology for the zero pressure gradient flat plate case were identical to those discussed in \citet{McConkeySciDataPaper2021}.

It should be noted that the predictions shown for the TBNN/KCNN use $S^\theta_{ij}$ for predicting the linear part of the anisotropy tensor. The nature of the proposed TBNN training process is to utilize $S^\theta_{ij}$ during training, so that the remaining non-linear part can be extracted during injection. 

\subsubsection{Flat plate (TBNN)}\label{sec:flat_plate}

\begin{figure}
    \centering
    \includegraphics{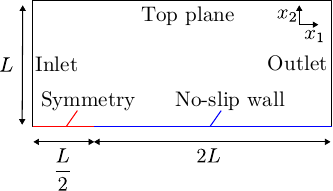}
    \caption{Computational domain for the zero pressure gradient flat plate boundary layer case.}
    \label{fig:flat_plate_domain}
\end{figure}

This case features a developing turbulent boundary layer on a flat plate with zero pressure gradient, based on the NASA ``2DZP" validation case~\citep{NASAturbmodel}. Figure~\ref{fig:flat_plate_domain} shows the domain for the flat plate case. The NASA-provided meshes are sufficient to resolve the viscous sublayer region, with a total number of cells $ N \approx 200,000$. However, this mesh was further refined to increase the number of solution data points available for training and testing. The goal of this case is to learn how the anisotropy tensor evolves in a turbulent boundary layer, therefore substantial mesh refinement was required to generate data points in this region. The total number of cells in the mesh is $N=4,673,130$. The plate-length Reynolds number is $Re_L= 5(10)^6$ to match the NASA reference data. The reference data for this case consists of a series of wall-normal profiles, for various $Re_\theta$, defined as 

\begin{equation}\label{eq:Re_theta}
    Re_\theta = \frac{U_\infty \theta }{\nu} \ ,
\end{equation}
where the momentum thickness $\theta$ is given as 

\begin{equation}
    \theta = \int_{0}^\infty \frac{U_{1}}{U_\infty} \left(1-\frac{U_1}{U_\infty}\right) d x_2 \ ,
\end{equation}
and $U_\infty$ is the free-stream velocity. The following boundary conditions and fluid properties are used for the domain in figure~\ref{fig:flat_plate_domain}. At the inlet boundary, $U_j$, $k$ , and $\omega$ are uniform: $U_j = (69.4, 0, 0) \text{ m/s}$, $k=1.08(10)^{-3} \ \text{m}^2/\text{s}^2$, $\omega = 8675 \ \text{s}^{-1}$, and $P$ is zero normal gradient. At the outlet, $P$ is zero, and all other variables are zero normal gradient. At the symmetry plane, all variables are zero normal gradient, and normal velocity is zero. At the top plane, all flow variables are zero normal gradient. At the no-slip wall, $U_j=0$, $k=0$, $\omega=6\nu/\beta_1 y^2$ (here, $\beta_1=0.075$), and $P$ is zero normal gradient. A kinematic viscosity of $\nu = 1.388(10)^{-5} \ \text{m}/\text{s}^2$ was used.

DNS reference data for a developing turbulent boundary layer comes from \citet{Schlatter2010}. This dataset contains a variety of turbulent boundary layer profiles, at various $Re_\theta$ shown in table~\ref{tbl:datasets}. As discussed in Section~\ref{sec:datasets}, the TBNN model was trained on various $Re_\theta$, with $Re_\theta=3,630$ serving as a hold-out test case. The results shown in this section are for this hold-out test case.

\begin{figure}
    \centering
    \includegraphics{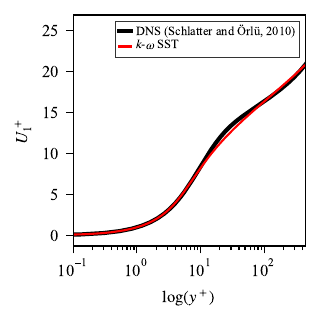}
    \caption[Streamwise velocity profile in the $Re_\theta=3630$ boundary layer predicted by the $k$-$\omega$ SST model, compared to the reference DNS data from \citet{Schlatter2010}.]{Streamwise velocity profile in the $Re_\theta=3630$ boundary layer predicted by the $k$-$\omega$ SST model, compared to the reference data from \citet{Schlatter2010}. Here, $U^+_1 = U^{}_1/u_\tau$, where $u_\tau=\sqrt{\tau_w/\rho}$, is the friction velocity, and $\tau_w$ is the wall shear stress. A density of $1$ kg/m$^3$ was used to be consistent with OpenFOAM's kinematic units. The wall-normal coordinate is $y^+=x_2u_\tau/\nu$.}
    \label{fig:BL_RANS}
\end{figure}

\begin{figure}
    \centering
    \includegraphics{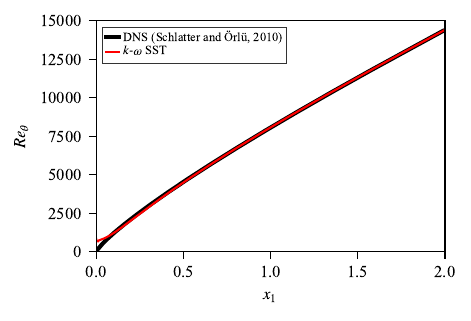}
    \caption{Momentum thickness Reynolds number growth along the flat plate as predicted by the $k$-$\omega$ SST model, and the reference DNS data from \citet{Schlatter2010}.}
    \label{fig:Retheta_RANS}
\end{figure}

The baseline $k$-$\omega$ SST model performs well for the test case flow in terms of predicting the mean velocity profiles. Figure~\ref{fig:BL_RANS} shows a sample mean velocity profile predicted by the baseline $k$-$\omega$ SST model, and figure~\ref{fig:Retheta_RANS} shows the predicted evolution of $Re_\theta$ along the plate. The excellent performance of the $k$-$\omega$ SST model demonstrated by Figures~\ref{fig:BL_RANS} and~\ref{fig:Retheta_RANS} is expected. This case features a fully attached boundary layer with zero pressure gradient, which is one of the fundamental calibration scenarios for RANS models. The zero pressure gradient turbulent boundary layer is considered a ``solved problem" for modern RANS models~\citep{Spalart2023}.

\begin{figure}
\centering
\begin{subfigure}[b]{\textwidth}
    \centering
    \includegraphics[width=0.55\textwidth]{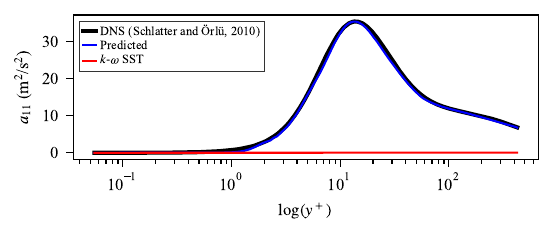}
    \caption{}
    \label{fig:flatplate_predictions_a11}
\end{subfigure}
\\
%\hfill
\begin{subfigure}[b]{\textwidth}
    \centering
    \includegraphics[width=0.55\textwidth]{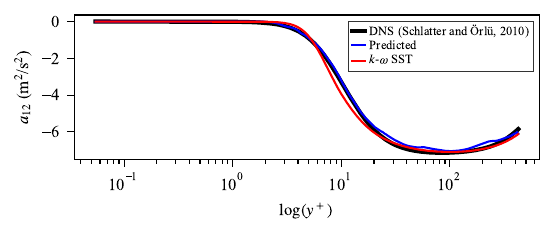}
    \caption{}
    \label{fig:flatplate_predictions_a12}
\end{subfigure}
\\
%\hfill
\begin{subfigure}[b]{\textwidth}
    \centering
    \includegraphics[width=0.55\textwidth]{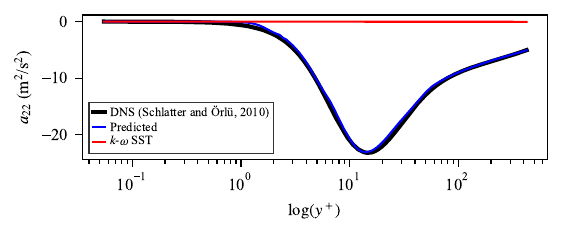}
    \caption{}
    \label{fig:flatplate_predictions_a22}
\end{subfigure}
\\
%\hfill
\begin{subfigure}[b]{\textwidth}
    \centering
    \includegraphics[width=0.55\textwidth]{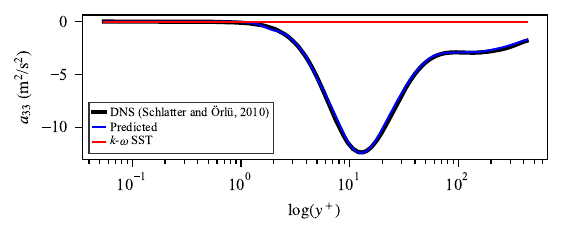}
    \caption{}
    \label{fig:flatplate_predictions_a33}
\end{subfigure}
%\hfill
\caption[Evolution of  $a_{11}$, $a_{12}$,  $a_{22}$, and $a_{33}$ in the $Re_\theta = 3630$ boundary layer, as predicted by the DNS data from \citet{Schlatter2010}, the TBNN/KCNN model, and the baseline $k$-$\omega$ SST model.]{Evolution of (a) $a_{11}$, (b) $a_{12}$, (c) $a_{22}$, and (d) $a_{33}$ in the $Re_\theta = 3630$ boundary layer, as predicted by the DNS data from \citet{Schlatter2010}, the TBNN/KCNN model, and the baseline $k$-$\omega$ SST model.}
\label{fig:flatplate_predictions}
\end{figure}

While the mean velocity profile is predicted well, figure~\ref{fig:flatplate_predictions} shows that the evolution of the near-wall anisotropy tensor is not predicted well. For this reason, the model architecture discussed in Section~\ref{sec:NN_architecture} was used to correct the anisotropy tensor in the near-wall region. This test also aims to determine whether the input feature set is sufficiently expressive to enable predicting the evolution of the anisotropy tensor within a boundary layer. Figure~\ref{fig:flatplate_predictions} shows the wall-normal profiles of various anisotropy tensor components predicted by the $k$-$\omega$ SST model, the ML-augmented $k$-$\omega$ SST model, and the reference DNS simulation for the hold-out test case. As discussed in Section~\ref{sec:datasets}, these models were trained on flat plate data at various values of $Re_\theta$. The results in figure~\ref{fig:flatplate_predictions} are designed to test these models on input features from an unseen boundary layer profile, to determine whether the learned anisotropy mapping was generalizable.

\begin{figure}
\centering
\begin{subfigure}[b]{\textwidth}
    \centering
    \includegraphics[width=0.55\textwidth]{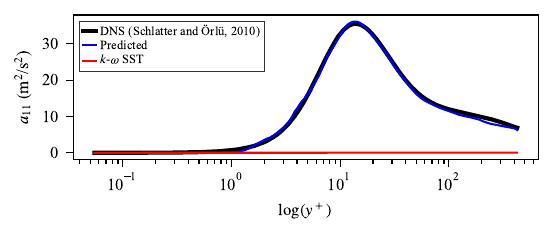}
    \caption{}
    \label{fig:kan_flatplate_predictions_a11}
\end{subfigure}
\\
%\hfill
\begin{subfigure}[b]{\textwidth}
    \centering
    \includegraphics[width=0.55\textwidth]{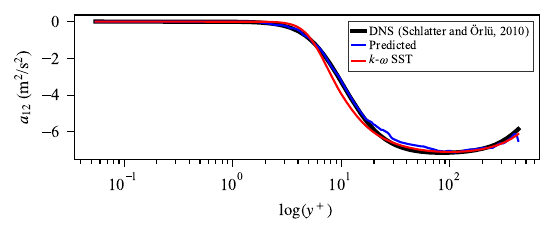}
    \caption{}
    \label{fig:kan_flatplate_predictions_a12}
\end{subfigure}
\\
%\hfill
\begin{subfigure}[b]{\textwidth}
    \centering
    \includegraphics[width=0.55\textwidth]{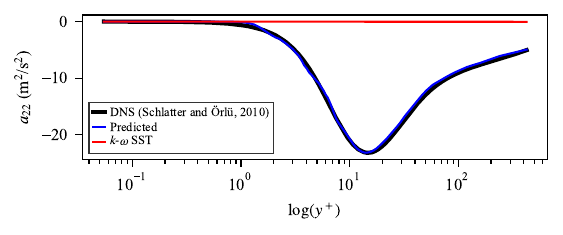}
    \caption{}
    \label{fig:kan_flatplate_predictions_a22}
\end{subfigure}
\\
%\hfill
\begin{subfigure}[b]{\textwidth}
    \centering
    \includegraphics[width=0.55\textwidth]{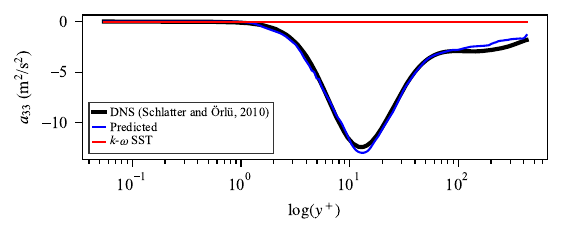}
    \caption{}
    \label{fig:kan_flatplate_predictions_a33}
\end{subfigure}
\caption[Anisotropy tensor component predictions in the flat plate boundary layer case.]{Evolution of (a) $a_{11}$, (b) $a_{12}$, (c) $a_{22}$, and (d) $a_{33}$ in the $Re_\theta = 3630$ boundary layer, as predicted by the DNS data from \citet{Schlatter2010}, the TBKAN/KCNN model, and the baseline $k$-$\omega$ SST model.}
\label{fig:kan_flatplate_predictions}
\end{figure}

Figure~\ref{fig:flatplate_predictions_a11} shows the predicted evolution of $a_{11}$ in the turbulent boundary layer. The baseline $k$-$\omega$ SST model predicts $a_{11}=0$, since $\partial U_1 /\partial x_1 \approx 0$ in the boundary layer. However, the DNS data clearly shows that $a_{11}$ is non-zero in the boundary layer. The TBNN/KCNN model combination is able to correct the the $a_{11}$ term to a high degree of accuracy in the boundary layer on this test case, indicating that the anisotropy mapping for the $a_{11}$ component generalizes well. Similar evolutions of $a_{22}$ (figure~\ref{fig:flatplate_predictions_a22}) and $a_{33}$ (figure~\ref{fig:flatplate_predictions_a33}) are observed in the DNS data. Again, the $k$-$\omega$ SST model predicts $a_{11}=a_{22}=a_{33}=0$, which is not physically correct. The TBNN/KCNN models are able to correct the baseline prediction to a high degree of accuracy on this unseen boundary layer profile.

Figure~\ref{fig:flatplate_predictions_a12} shows the predicted evolution of $a_{12}$. The baseline RANS model predicts the evolution of $a_{12}$ well, and this is likely the reason that the mean velocity profile of $U_{1}$ is predicted well (see figure~\ref{fig:BL_RANS}). While the TBNN/KCNN is not needed to correct this off-diagonal component, it is able to correct minor inaccuracies in the $k$-$\omega$ SST model predictions in the buffer region ($5 \leq y^+ \leq 30$). Nevertheless, the baseline $k$-$\omega$ SST model achieves a satisfactory accuracy level for this flow. As discussed, this is expected, given that low Reynolds number RANS models are able to predict a zero pressure gradient boundary layer with a high degree of accuracy. Figure~\ref{fig:flatplate_predictions} demonstrates that this performance is the result of an accurate prediction of $a_{12}$ by the $k$-$\omega$ SST model.

%inserting the TBKAN results here as they are to be shifted
\subsubsection{Flat plate (TBKAN)}

The predicted evolution of the various components of the anisotropy tensor for the flat plate case obtained using the TBKAN/KCNN model combination is displayed in figure~\ref{fig:kan_flatplate_predictions_a11}--\ref{fig:kan_flatplate_predictions_a33} for the hold-out test case $Re_\theta = 3,630$. These predictions are compared to both the baseline $k$-$\omega$ SST model and DNS data from \citet{Schlatter2010}. For this test case, the predictive accuracy of the TBKAN/KCNN model combination for the anisotropy tensor components is compared to that of the baseline model.

The predicted evolution of \(a_{11}\) is shown in figure~\ref{fig:kan_flatplate_predictions_a11}. Unlike the $k$-$\omega$ SST model, which predicts \(a_{11}=0\) due to its isotropic stress assumptions, the TBKAN captures the non-zero nature of \(a_{11}\) in the turbulent boundary layer. The TBKAN aligns closely with the DNS data, indicating its ability to generalize and represent anisotropy accurately, particularly for components where the baseline model is limited.

Figure~\ref{fig:kan_flatplate_predictions_a12} presents the predicted evolution of \(a_{12}\). The baseline $k$-$\omega$ SST model predicts this off-diagonal component with reasonable accuracy. However, the TBKAN/KCNN exhibits better conformance with the reference DNS data in the buffer region ($5 \leq y^+ \leq 30$), correcting minor deviations in the predictions of this quantity provided by the baseline $k$-$\omega$ SST model. 

Figures~\ref{fig:kan_flatplate_predictions_a22} and \ref{fig:kan_flatplate_predictions_a33} display the predictions of \(a_{22}\) and \(a_{33}\), respectively. While the $k$-$\omega$ SST model predicts a value of zero from these two anisotropy stress components, the TBKAN/KCNN captures their evolution with good accuracy compared to the DNS data. Overall, TBKAN/KCNN enhances the predictive accuracy of the anisotropic stress components across the boundary layer. While \(a_{12}\) predictions are marginally improved compared to the baseline model, TBKAN/KCNN shows substantial improvements for the normal stress components. Furthermore, a visual perusal of Figures \ref{fig:flatplate_predictions} and \ref{fig:kan_flatplate_predictions} shows that the conformance of the predictions of the anisotropy tensor components with the reference DNS data obtained with TBKAN/KCNN is marginally worse than that obtained with TBNN/KCNN. However, it is noted that the TBKAN is simpler than TBNN in this case in the sense that the former network used only one hidden layer with 9 nodes (where the information from the edges encoded in the B-splines are simply accumulated), whereas the latter network used four hidden layers consisting of 20 nodes each (where the information embodied by the linear weights in the edges is transformed by the nonlinear activation function).

\subsubsection{Square duct}\label{sec:square_duct}
Turbulent flow through a square duct is a challenging case for RANS models, since linear eddy viscosity models cannot predict the secondary flows that occur in the cross-sectional plane. The goal of the square duct test case is to determine whether the proposed closure framework could enable the $k$-$\omega$ SST model to predict these Prandtl secondary flows~\citep{Nikitin2021}. The square duct DNS dataset generated by \citet{Pinelli2010} was used as reference data, and the RANS data from \citet{McConkeySciDataPaper2021} was used.

\begin{figure}
\centering
\begin{subfigure}[b]{0.6\textwidth}
    \centering
    \includegraphics{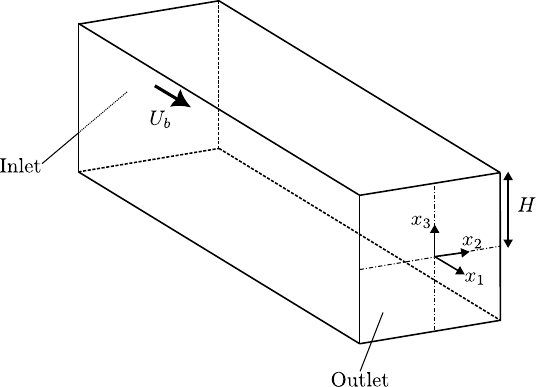}
    \caption{}
    \label{fig:duct_domain}
\end{subfigure}
\hfill
\begin{subfigure}[b]{0.3\textwidth}
    \centering
    \includegraphics[width=\textwidth]{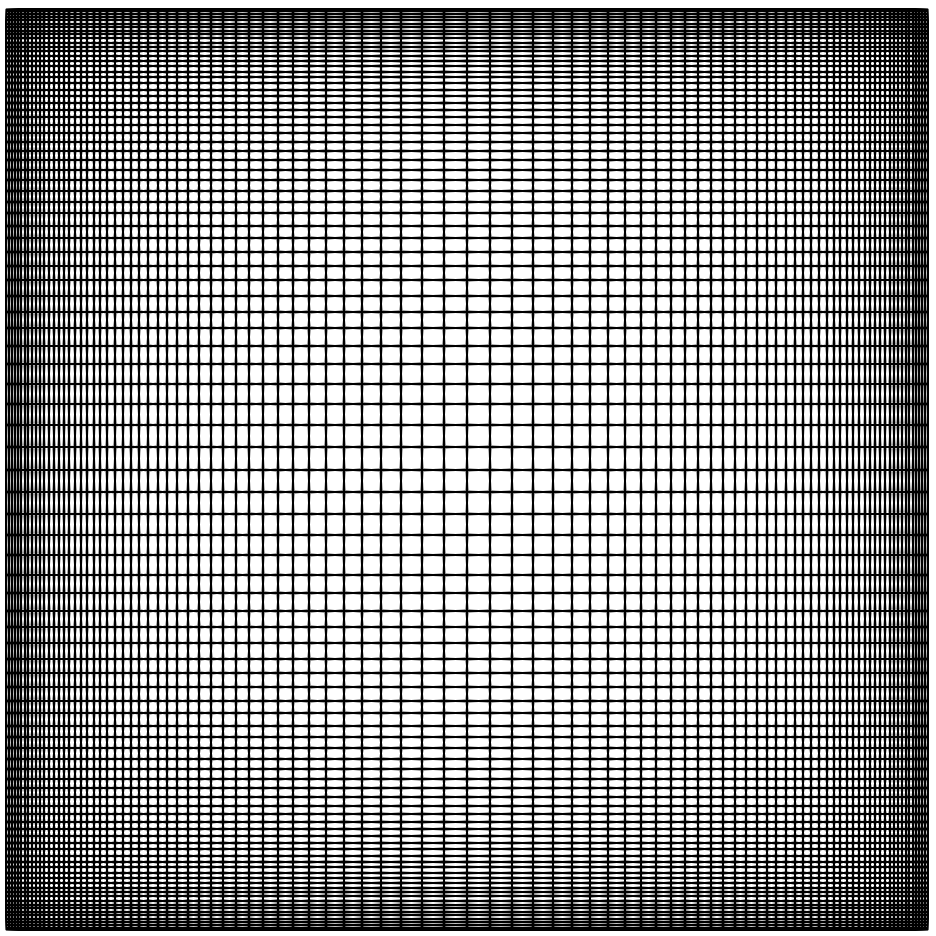}
    \caption{}
    \label{fig:duct_mesh}
\end{subfigure}
\hfill
\caption[Computational domain and RANS mesh for the square duct flow.]{Computational (a) domain and (b) RANS mesh for the square duct flow.}\label{fig:square_duct}
\end{figure}

Figure~\ref{fig:square_duct} shows the computational setup and mesh for the square duct case. The mesh is designed to achieve $y^+\leq 1$ for all square duct cases. As discussed in Section~\ref{sec:datasets}, the duct half-height $H$ Reynolds number varies between cases, calculated by:
\begin{equation}
    Re_H = \frac{U_b H}{\nu} \ ,
\end{equation}
where $U_b$ is the mean (bulk) cross-sectional velocity. A kinematic viscosity of $\nu=5(10)^{-6} $ m$^2$/s was used for all square duct cases. With the geometry fixed as shown in figure~\ref{fig:square_duct}, the bulk velocity was adjusted to vary the Reynolds number. More details on the computational setup for the square duct case are provided by \citet{McConkeySciDataPaper2021}. The boundary conditions are periodic at the inlet/outlet, and no-slip walls were applied along the sides of the duct. As discussed in Section~\ref{sec:datasets}, the modified TBNN was trained on several values of $Re_H$, with $Re_H = 2,000$ serving as a hold-out test case. 

\begin{figure}
    \centering
    \includegraphics[width=\textwidth]{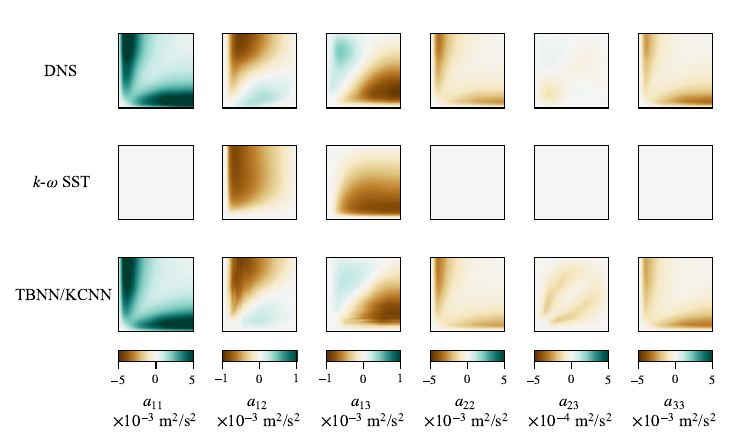}
    \caption[Components of $a_{ij}$ predicted by the DNS data from \citet{Pinelli2010}, the $k$-$\omega$ SST model, and the TBNN/KCNN model from the present investigation.]{Components of $a_{ij}$ predicted by the DNS data from \citet{Pinelli2010}, the $k$-$\omega$ SST model, and the TBNN/KCNN model from the present investigation. Shown here are contours of $a_{ij}$ components in the lower left $x_2\leq0, \ x_3\leq0$ quadrant.}
    \label{fig:duct_a}
\end{figure}

Figure~\ref{fig:duct_a} shows the components of the anisotropy tensor $a_{ij}$ predicted by RANS, DNS, and the TBNN/KCNN models for the square duct test case. The $k$-$\omega$ SST model is a linear eddy viscosity model, and therefore predicts zero $a_{ij}$ where $S_{ij}$ is zero. Figure~\ref{fig:duct_a} shows that $a_{11}$, $a_{22}$, $a_{23}$, and $a_{33}$ are all non-zero in the duct, and that the $k$-$\omega$ SST model is unable to capture this behavior. The TBNN/KCNN models predict an accurate evolution of almost all anisotropy tensor components across the duct cross-section (viz. $a_{11}$, $a_{12}$, $a_{13}$, $a_{22}$, and $a_{33}$ are all predicted well on this test case). The anisotropy tensor component $a_{23}$ is not predicted well, likely because it is at least an order of magnitude smaller than the other components, and therefore errors in $a_{23}$ are not penalized as heavily in the loss function.

\begin{figure}
\centering
\begin{subfigure}[b]{0.45\textwidth}
    \centering
    \includegraphics{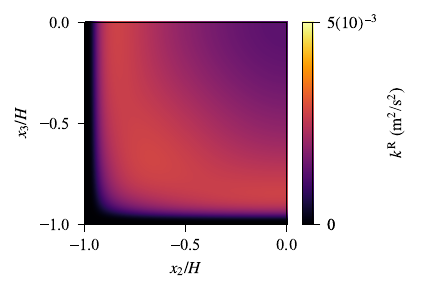}
    \caption{}
    \label{fig:duct_k_RANS}
\end{subfigure}
\hfill
\begin{subfigure}[b]{0.45\textwidth}
    \centering
    \includegraphics{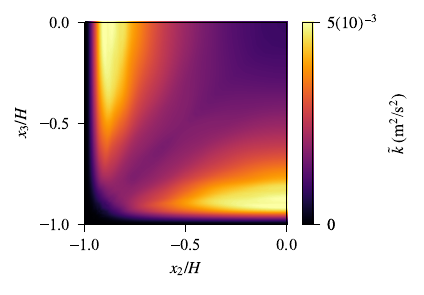}
    \caption{}
    \label{fig:duct_k_pred}
\end{subfigure}
\hfill
\begin{subfigure}[b]{0.45\textwidth}
    \centering
    \includegraphics{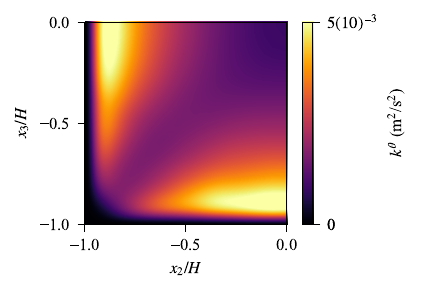}
    \caption{}
    \label{fig:duct_k_DNS}
\end{subfigure}
\hfill
\caption[Contours of turbulent kinetic energy $k$ predicted by the $k$-$\omega$ SST model, the KCNN in the present investigation, and the DNS data from \citet{Pinelli2010}.]{Contours of turbulent kinetic energy $k$ predicted by (a) the $k$-$\omega$ SST model, (b) the KCNN in the present investigation, and (c) the DNS data from \citet{Pinelli2010}. Shown here is the lower left $x_2\leq0, \ x_3\leq0$ quadrant.}\label{fig:duct_k}
\end{figure}

Figure~\ref{fig:duct_k} shows the turbulent kinetic energy $k$ after being corrected by the KCNN model for the square duct test case. Accurate prediction of $k$ is critical to an accurate estimate of $a_{ij}$, since $a_{ij}=2kb_{ij}$. The $k$-$\omega$ SST model generally under-predicts $k$. After correction via the KCNN, the $k$ field is predicted well compared to the DNS data. The primary feature in the $k$ field that is absent from the $k$-$\omega$ SST prediction is the high-$k$ region along the side walls of the duct. The KCNN introduces a correction to the baseline RANS field, and is able to predict this high-$k$ region.

Ultimately, it is the goal of the proposed framework to improve the estimated mean fields in the RANS simulation. To determine whether the corrected closure term would produce corrected mean velocity fields, the predicted $\tilde a^\perp_{ij}$ was injected into the RANS momentum equation as shown in Equation~\ref{eq:RANS_new}. The momentum and continuity equations converged around the fixed $\tilde a^\perp_{ij}$ until numerical convergence was achieved. In OpenFOAM v2212, a modified version of the PIMPLE solver was implemented for the purpose of this injection. The PIMPLE solver was used to incorporate an unsteady term into the system of equations during iteration, to promote stability. Though this unsteady term affects the solution during convergence, the simulation ultimately achieved a steady-state condition, thereby reducing this unsteady term to zero.

\begin{figure}
\centering
\begin{subfigure}[b]{0.51\textwidth}
    \centering
    \includegraphics{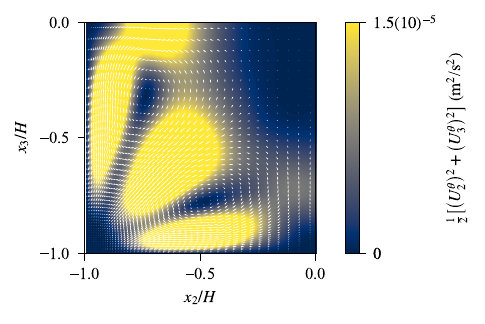}
    \caption{}
    \label{fig:duct_vectors_DNS}
\end{subfigure}
\hfill
\begin{subfigure}[b]{0.51\textwidth}
    \centering
    \includegraphics{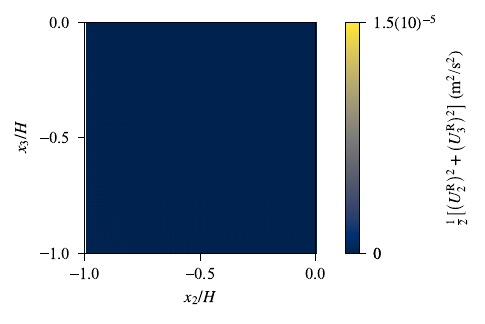}
    \caption{}
    \label{fig:duct_vectors_rans}
\end{subfigure}
\hfill
\begin{subfigure}[b]{0.51\textwidth}
    \centering
    \includegraphics{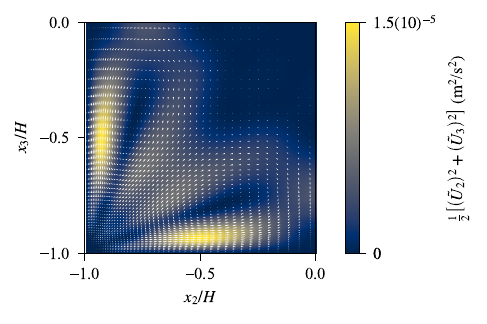}
    \caption{}
    \label{fig:duct_vectors_pred}
\end{subfigure}
\hfill

\caption[Velocity vectors and in-plane kinetic energy predicted by the DNS data from \citet{Pinelli2010} and injecting the TBNN/KCNN predictions into the RANS momentum equations. ]{Velocity vectors and in-plane kinetic energy predicted by (a) the DNS data from \citet{Pinelli2010}, (b) the $k$-$\omega$ SST model, and (c) injecting the TBNN/KCNN predictions into the RANS momentum equations. Shown here is the lower left $x_2\leq0, \ x_3\leq0$ quadrant.} \label{fig:duct_vectors}
\end{figure}

Figure~\ref{fig:duct_vectors} shows that the TBNN/KCNN model is able to produce secondary flows after injecting $\tilde a^\perp_{ij}$ into the momentum equation. This \textit{a posteriori} prediction of the mean field is ultimately the main prediction of interest for a ML-augmented RANS closure framework. Whereas the original $k$-$\omega$ SST model does not predict formation of any secondary flows in the duct, figure~\ref{fig:duct_vectors} shows that the ML-augmented $k$-$\omega$ SST model predicts corner vortices. However, the in-plane kinetic energy is generally underpredicted by the ML-augmented RANS model.

\begin{figure}
\centering
\begin{subfigure}[b]{0.45\textwidth}
    \centering
    \includegraphics{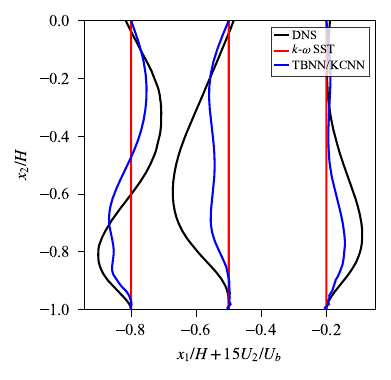}
    \caption{}
    \label{fig:duct_profiles_U2}
\end{subfigure}
\hfill
\begin{subfigure}[b]{0.45\textwidth}
    \centering
    \includegraphics{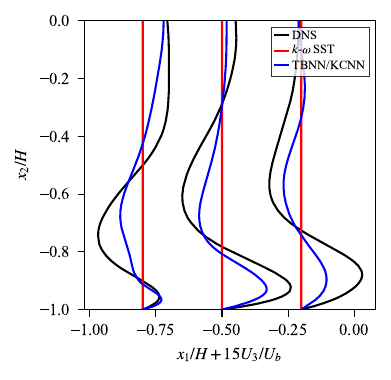}
    \caption{}
    \label{fig:duct_profiles_U3}
\end{subfigure}
\hfill

\caption[Profiles of the in-plane velocity components $U_2$ and $U_3$, as predicted by the $k$-$\omega$ SST model, the injected TBNN/KCNN predictions, and the DNS data from \citet{Pinelli2010}.] {Profiles of the in-plane velocity components (a) $U_2$ and (b) $U_3$, as predicted by the $k$-$\omega$ SST model, the injected TBNN/KCNN predictions, and the DNS data from \citet{Pinelli2010}. Shown here is the lower left $x_2\leq0, \ x_3\leq0$ quadrant.} 
\label{fig:duct_velocity_profiles}
\end{figure}

To further examine the ability of the ML-augmented $k$-$\omega$ SST model to predict secondary flows in the duct test case, profiles of $U_2$ and $U_3$ are plotted in figure~\ref{fig:duct_velocity_profiles}. While the ML-augmented model is able to produce this non-linear feature, the corner vortex strengths are reduced compared to the reference DNS data. Both the $U_2$ and $U_3$ components are under-predicted. Nevertheless, the $k$-$\omega$ SST model (which predicts $U_2=U_3=0$) has clearly been improved via a ML-augmented correction to the closure term in the momentum equation. From figure~\ref{fig:duct_a}, it would appear that the good prediction of the normal stress anisotropy (the primary mechanism responsible for the streamwise vorticity determined by $U_2$ and $U_3$) should provide good predictions of the streamwise vorticity. However, once the secondary flow is set in motion by this normal stress anisotropy, it is the secondary (rather than primary) shear stress component $a_{23}$ (generated by the presence of the secondary flow itself) that is required to maintain this flow and from figure~\ref{fig:duct_a}, this secondary component of the shear stress is not well predicted. Therefore, the underprediction of $U_2$ and $U_3$ is likely due to inaccurate prediction of $a_{23}$.

\subsubsection{Rectangular duct test case}

Duct aspect ratio has an influence on the in-plane kinetic energy $\frac{1}{2}\left(U_2^2+U_3^2\right)$ and behaviour of the corner vortices~\citep{Vinuesa_duct2,Vinuesa_duct4,Vinuesa_duct3}. It was of interest to examine the ability of a model trained on square ducts to generalize to a rectangular duct at similar Reynolds number. The TBNN/KCNN models were applied to a duct with aspect ratio 3 and $Re_H\approx 2600$, matching the DNS simulation by ~\citet{Vinuesa_duct1}. This generalization test was carried out in the same manner as the $Re_H=2000$ hold-out test case (i.e, by making a predictive correction to the closure terms in the $k$-$\omega$ SST turbulence model, and injecting them back into the momentum equation). 

\begin{figure}
\centering
\begin{subfigure}[b]{0.45\textwidth}
    \centering
    \includegraphics{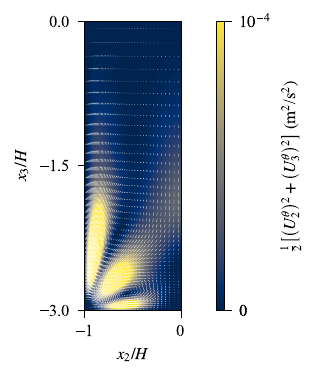}
    \caption{}
    \label{fig:ar3_vectors_DNS}
\end{subfigure}
\hfill
\begin{subfigure}[b]{0.45\textwidth}
    \centering
    \includegraphics{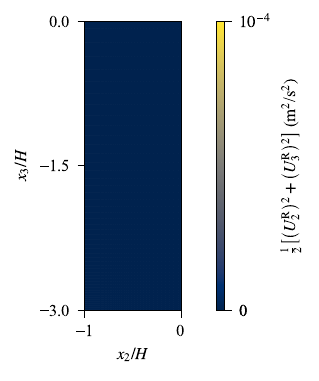}
    \caption{}
    \label{fig:ar3_vectors_rans}
\end{subfigure}
\hfill
\begin{subfigure}[b]{0.45\textwidth}
    \centering
    \includegraphics{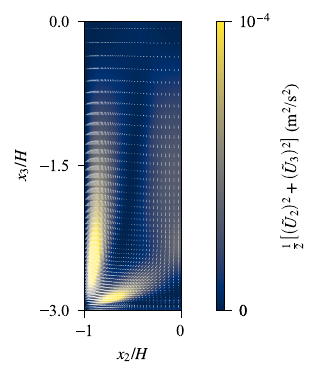}
    \caption{}
    \label{fig:ar3_vectors_pred}
\end{subfigure}
\hfill

\caption[Velocity vectors and in-plane kinetic energy predicted by the DNS data from \citet{Vinuesa_duct} and injecting the TBNN/KCNN predictions into the RANS momentum equations. ]{Velocity vectors and in-plane kinetic energy predicted by (a) the DNS data from \citet{Pinelli2010}, (b) the $k$-$\omega$ SST model, and (c) injecting the TBNN/KCNN predictions into the RANS momentum equations. Shown here is the lower left $x_2\leq0, \ x_3\leq0$ quadrant.} \label{fig:ar3}
\end{figure}

Figure~\ref{fig:ar3} compares the in-plane kinetic energy and velocity vector field predicted by the baseline $k$-$\omega$ SST model, and the TBNN/KCNN augmented $k$-$\omega$ SST model. Examining figure~\ref{fig:ar3} shows that ML-based correction enables the augmented SST model to predict secondary flows. However, in the same manner as the square duct test case, the in-plane kinetic energy ins underpredicted. Additionally, the stength of the dominant corner vortex is over-predicted by the ML-augmented model, while the  stength of the smaller corner vortex is significantly underpredicted. The magnitude of the in-plane kinetic energy is similar, however the peak locations of this field are also shifted due to a mismatch in predicted corner vortex shape. Nevertheless, the ML-augmented model shows improvement compared to the $k$-$\omega$ SST model, which fails to predict any corner vortices. In order to promote better generalization to rectangular geometries, a more extensive training dataset consisting of rectangular ducts would need to be used to train the models.

\subsubsection{Periodic hills}\label{sec:periodic_hills}
Flow over periodic hills is used as a popular benchmark case for turbulence modelling given the challenging physics of boundary layer separation in an adverse pressure gradient, reattachment along the bottom wall, and acceleration of the flow before reentering the domain. For the purpose of data-driven turbulence modelling, a variety of periodic hills data has been made available. In this study, we use the $Re_H=5,600$ configuration, simulated using DNS by \citet{Xiao2020}. Xiao et al.'s data was included in \citet{McConkeySciDataPaper2021}, which is the primary data source for this study.

\begin{figure}
    \centering
    \includegraphics[width=\textwidth]{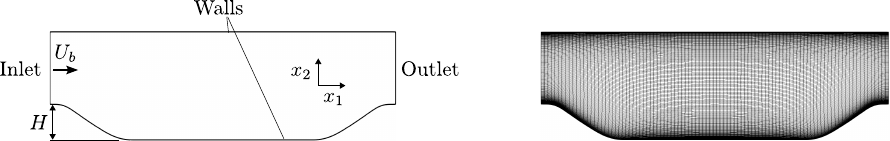}
    \caption{Computational domain and mesh for the periodic hills case, with $\alpha=1.2$.}
    \label{fig:periodic_hills}
\end{figure}
The geometry and mesh for the periodic hills case are shown in figure~\ref{fig:periodic_hills}. For all periodic hills cases, the hill height-based Reynolds number is 5,600, calculated by

\begin{equation}
    Re_H = \frac{U_b H}{\nu} \ ,
\end{equation}
where $U_b$ is the bulk (mean) velocity at the domain inlet. The hill geometry is varied between cases, based on the hill steepness $\alpha$. Further details on the computational setup for the baseline RANS periodic hills simulations are provided in \citet{McConkeySciDataPaper2021}. The boundary conditions are periodic at the inlet/outlet, and no-slip walls at the top and bottom of the domain were imposed. As discussed in Section~\ref{sec:datasets}, the TBNN and KCNN models were trained on several hill steepness values, with $\alpha=1.2$ being used as a hold-out test set. 

\begin{figure}
    \centering
    \includegraphics[width=\textwidth]{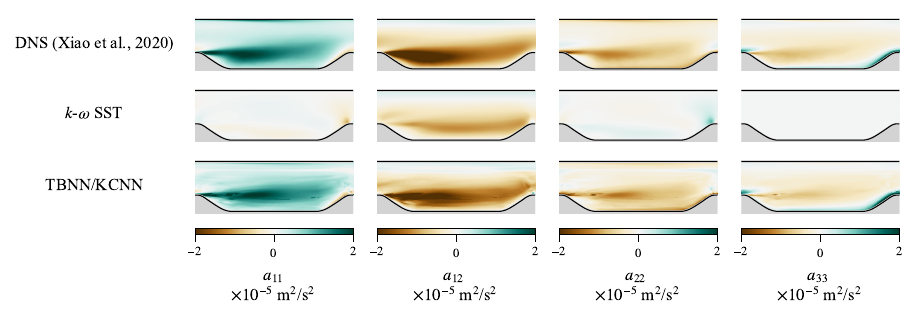}
    \caption{Contours of non-zero $a_{ij}$ components predicted by the DNS data from \citet{Xiao2020}, the $k$-$\omega$ SST model, and the  TBNN/KCNN model predictions from the present investigation.}
    \label{fig:periodic_hills_a}
\end{figure}

Figure~\ref{fig:periodic_hills_a} shows the components of the anisotropy tensor $a_{ij}$ predicted by RANS ($k$-$\omega$ SST), DNS, and the ML-augmented RANS simulation. The improvement in all components of $a_{ij}$ is clear. The baseline $k$-$\omega$ SST under-predicts all components, with severe under-prediction of $a_{11}$, $a_{22}$, and $a_{33}$. The $a_{12}$ prediction by the $k$-$\omega$ SST showcases similar trends to the DNS data, but the overall magnitude is lower. However, after correction, key features of all $a_{ij}$ fields are captured when the TBNN/KCNN augment the $k$-$\omega$ SST model. In particular, the higher magnitudes of the diagonal $a_{ij}$ (normal stress) components are captured by the augmented model.

\begin{figure}
    \centering
    \includegraphics[width=\textwidth]{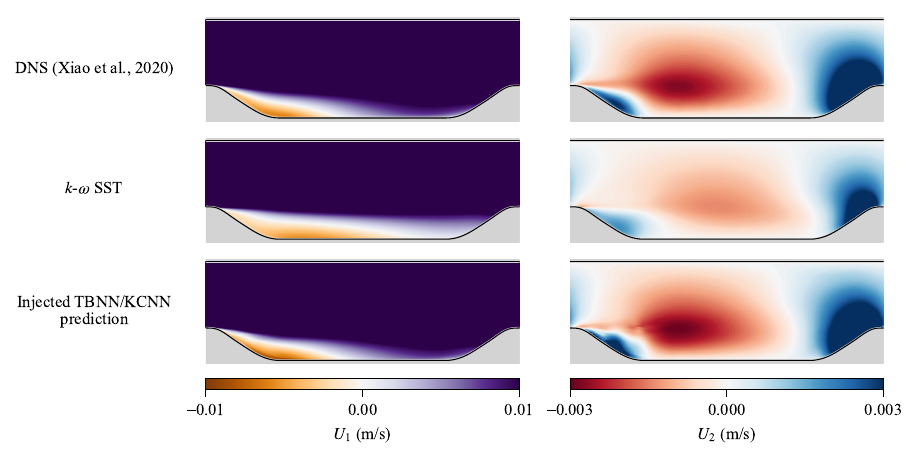}
    \caption{Contours of $U_1$ and $U_1$ predicted by the DNS data from \citet{Xiao2020}, the $k$-$\omega$ SST model, and the injected TBNN/KCNN model predictions from the present investigation.}
    \label{fig:periodic_hills_U}
\end{figure}

\begin{figure}
    \centering
    \includegraphics[width=\textwidth]{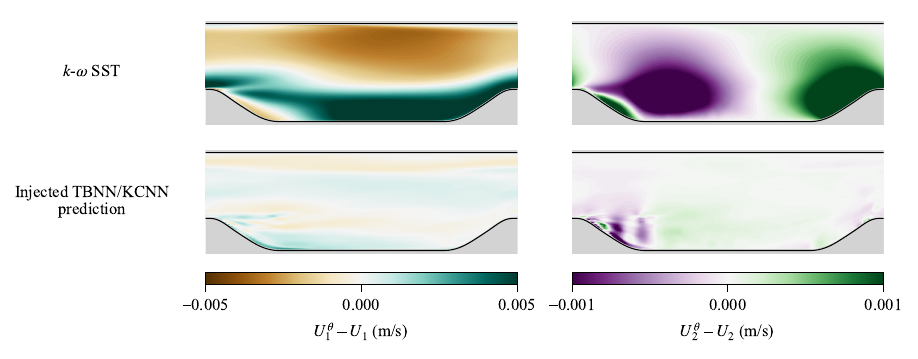}
    \caption{Contours of error in $U_1$ and $U_2$ predicted by the $k$-$\omega$ SST model and the injected TBNN/KCNN model predictions, as compared to the DNS data from \citet{Xiao2020}.}
    \label{fig:periodic_hills_U_errors}
\end{figure}

As was done for the square duct test case (Section~\ref{sec:square_duct}), a modified PIMPLE solver was used to inject the predicted $\tilde a_{ij}$ into the RANS momentum equation for the periodic hills test case. The numerical setup for the periodic hills injection was identical to the square duct case. Figure~\ref{fig:periodic_hills_U} compares the mean velocity fields before and after the corrected closure term is used within the RANS simulation. Figure~\ref{fig:periodic_hills_U_errors} compares the errors in the velocity components $U_1$ and $U_2$ estimated by the $k$-$\omega$ SST model, and the \textit{a posteriori} (post-injection) TBNN/KCNN-augmented SST model.

As seen in figure~\ref{fig:periodic_hills_U}, the primary feature of this flow is a recirculation zone which appears immediately after the left hill. The recirculation zone is most clearly visualized by examining the $U_1$ fields. The $k$-$\omega$ SST model over-predicts the size of this recirculation zone. After correction via injecting the TBNN/KCNN predictions, the recirculation zone size closely matches the DNS data. In the $U_2$ field, a region with $U_2<0$ is seen immediately above this recirculation region. The baseline $k$-$\omega$ SST model under-predicts the downward velocity here, leading to delayed reattachment, and a longer recirculation zone. After correction, the magnitude of $U_2$ in this shear layer more closely matches the DNS data. On the right hill, the upward acceleration of the flow under the favourable pressure gradient is also under-predicted by the $k$-$\omega$ SST. Here, the injected $\tilde a_{ij}$ is able to better capture the strength of this upward acceleration.

Figure~\ref{fig:periodic_hills_U_errors} more closely examines the improvements offered by augmenting the $k$-$\omega$ SST model via the TBNN/KCNN. It can be seen that the overall magnitudes of the errors in $U_1$ and $U_2$ are significantly reduced after injecting the TBNN/KCNN model predictions. In particular, error in $U_1$ is reduced in the reattachment region along the bottom wall, and the bulk flow above this region. Error in $U_2$ is reduced in the previously identified shear layer above the recirculation region, and the accelerating region before the outlet.

\subsection{Impact of realizability-informed training}\label{sec:realizability_results}
To determine the impact that including a realizability-informed penalty has on the training process, the closure term predictions for the three test cases were examined in greater detail. Two loss functions were used: one with $\alpha=0$ (representing no realizability penalty), and one with $\alpha=10^2$ (representing an exaggerated realizability penalty term in the loss function). The objective of this test was to determine whether including the realizability penalty during training promotes better generalization of the closure mapping to unseen flow variations.

All TBNN hyperparameters were fixed to those given in table~\ref{tbl:hyperparameters_ri}. Since the realizability-informed training procedure only applies to the TBNN, a perfect prediction of $k$ via the KCNN was assumed for calculating error in $\tilde a_{ij}$. 

Table~\ref{tbl:realizability_test} compares the mean-squared error in $\tilde a_{ij}$ on the hold-out test set, with and without realizability penalties being used in training. It should be noted that similar to the \textit{a priori} tests in Section~\ref{sec:results_ri}, the linear component used when visualizing $b_{ij}$ comes from $S^\theta _{ij}$, as is the configuration when training the TBNN. This linear component is the one used when training the TBNN, and therefore its use here provides the most fair assessment of how the proposed loss function promotes more physically realizable results.

\begin{table}
\centering
\caption{Comparison of mean squared error and number of non-realizable predictions when training with and without a realizability-informed loss function.}
\label{tbl:realizability_test}
\fontsize{10}{12}\selectfont
\renewcommand{\arraystretch}{1.25}
\begin{tabular}{cccc}
\hline
Dataset                         & $\alpha$ & MSE ($a_{ij}$)  & $\%$ non-realizable \\ \hline
\multirow{2}{*}{Flat plate}     & 0        & $3.5(10)^{-3}$  & $0.6\%$             \\
                                & $10^2$   & $7.4(10)^{-3}$  & $0\%$               \\ \hline
\multirow{2}{*}{Square duct}    & 0        & $2.2(10)^{-9}$  & $9.2\%$             \\
                                & $10^2$   & $2.6(10)^{-9}$  & $0.4\%$             \\ \hline
\multirow{2}{*}{Periodic hills} & 0        & $3.9(10)^{-13}$ & $1.7\%$             \\
                                & $10^2$   & $6.3(10)^{-13}$ & $0.3\%$             \\ \hline
\end{tabular}
\end{table}

As seen in table~\ref{tbl:realizability_test}, the realizability-informed loss function significantly reduces realizability violations on unseen flow variations. Even on hold-out test cases, the model is able to predict $b_{ij}$ without any realizability violations. In some cases, a small tradeoff in $a_{ij}$ occurs---this tradeoff is expected, as for some difficult points the realizability-informed loss function involves a tradeoff between error and realizability. However, in all cases, realizability-informed training also reduces the error in $b_{ij}$. This error reduction in $b_{ij}$ is expected, since all $b_{ij}$ reference data are realizable. In the case of predicting $b_{ij}$ accurately, the gradients of the realizability penalty $\mathcal{R}(b_{ij})$ further push the predictions towards an accurate prediction of $b_{ij}$, compared to a purely mean-squared error  gradient.

\begin{figure}
\centering
\begin{subfigure}[b]{0.24\textwidth}
    \centering
    \includegraphics{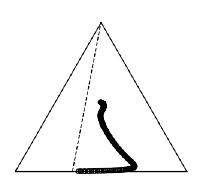}
    \caption{}
    \label{fig:fp_barycentric}
\end{subfigure}
\hfill
\begin{subfigure}[b]{0.24\textwidth}
    \centering
    \includegraphics{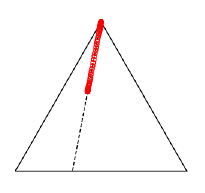}
    \caption{}
    \label{fig:duct_barycentric}
\end{subfigure}
\hfill
\begin{subfigure}[b]{0.24\textwidth}
    \centering
    \includegraphics{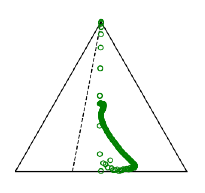}
    \caption{}
    \label{fig:phll_barycentric}
\end{subfigure}
\hfill
\begin{subfigure}[b]{0.24\textwidth}
    \centering
    \includegraphics{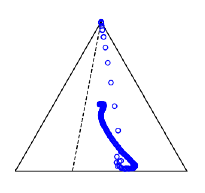}
    \caption{}
\end{subfigure}
\hfill
\centering
\begin{subfigure}[b]{0.24\textwidth}
    \centering
    \includegraphics{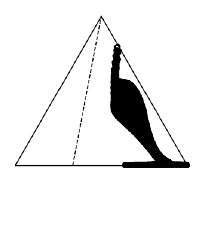}
    \caption{}
    \label{fig:fp_barycentric}
\end{subfigure}
\hfill
\begin{subfigure}[b]{0.24\textwidth}
    \centering
    \includegraphics{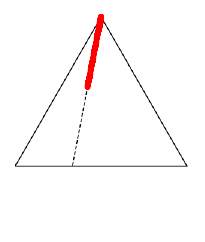}
    \caption{}
    \label{fig:duct_barycentric}
\end{subfigure}
\hfill
\begin{subfigure}[b]{0.24\textwidth}
    \centering
    \includegraphics{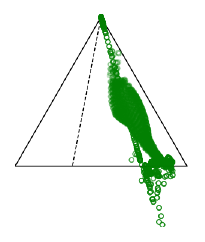}
    \caption{}
    \label{fig:phll_barycentric}
\end{subfigure}
\hfill
\begin{subfigure}[b]{0.24\textwidth}
    \centering
    \includegraphics{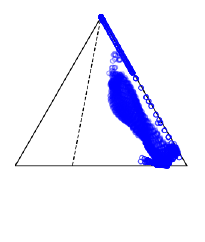}
    \caption{}
\end{subfigure}
\hfill

\centering
\begin{subfigure}[b]{0.24\textwidth}
    \centering
    \includegraphics{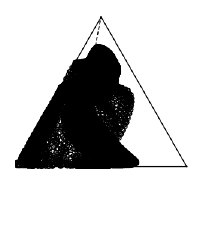}
    \caption{}
    \label{fig:fp_barycentric}
\end{subfigure}
\hfill
\begin{subfigure}[b]{0.24\textwidth}
    \centering
    \includegraphics{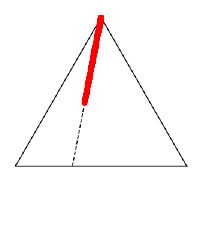}
    \caption{}
    \label{fig:duct_barycentric}
\end{subfigure}
\hfill
\begin{subfigure}[b]{0.24\textwidth}
    \centering
    \includegraphics{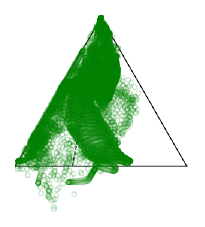}
    \caption{}
    \label{fig:phll_barycentric}
\end{subfigure}
\hfill
\begin{subfigure}[b]{0.24\textwidth}
    \centering
    \includegraphics{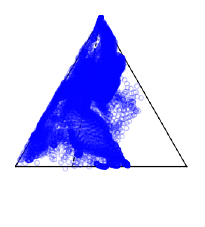}
    \caption{}
\end{subfigure}
\hfill

\begin{subfigure}[b]{0.5\textwidth}
    \centering
    \includegraphics{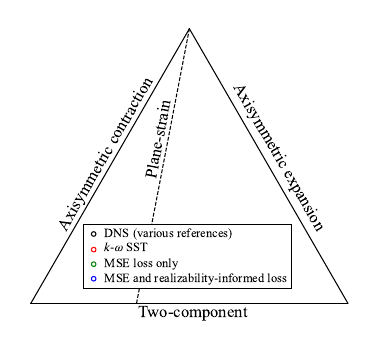}
    \caption{}
\end{subfigure}
\hfill
\caption[Projection of the predicted $b_{ij}$ from the DNS reference data (a, e, i), the $k$-$\omega$ SST model (b, f, j),  and the TBNN/KCNN with (d, h, l) and without (c, g, k) realizability-informed learning onto the barycentric triangle.]{Projection of the predicted $b_{ij}$ from the DNS reference data (a, e, i), the $k$-$\omega$ SST model (b, f, j),  and the TBNN/KCNN with (d, h, l) and without (c, g, k) realizability-informed learning onto the barycentric triangle. (a)--(d) show the flat plate data, (e)--(h) show the square duct data, and (i)--(l) show the periodic hills data. Points outside the triangle are not realizable.}
\label{fig:barycentric}
\end{figure}

To further visualize the realizability of the TBNN predictions, the barycentric map of \citet{Banerjee2007} was used. In the barycentric map, the eigenvalues of a given $b_{ij}$ are mapped into a triangle, the bounds of which represent the limiting realizable behaviors of turbulent fluctuations. This triangle is useful to spatially visualize realizability violations, and types of turbulent flow physics predicted by the baseline turbulence model, the ML-augmented turbulence model, and DNS. Further details on the construction of this mapping are given in \citet{Banerjee2007}, and \citet{Emory2014}.

Figure~\ref{fig:barycentric} compares the realizability of the predictions made by a model trained on only a mean-squared error loss function to a model trained on the realizability-informed loss function for the three flows in the present study. All predictions are for the hold-out test set for each flow, representing a generalization test. The goal of the realizability-informed loss function is to encode a preference for realizable predictions into the model when generalizing outside of the training dataset. Figure~\ref{fig:barycentric} shows a clear improvement in the realizability of the TBNN predictions. This visualization supports the results in table~\ref{tbl:realizability_test} in that a realizability-informed model has significantly lower realizability violations when predicting the anisotropy tensor on a new flow. Nearly all of the predictions from the realizability-informed model fall within the realizable boundaries, while the model trained only on mean-squared error predicts several realizability-violating results when generalizing to new cases of complex flows such as the duct and periodic hills cases. Also, the violations of physical realizability for the realizability-informed model (when they do occur) are not as severe (as measured from the magnitude of deviation outside the barycentric map) as those obtained from only a mean-squared error loss function. 

Figure~\ref{fig:barycentric} also shows that for all flows in the present study, the original $k$-$\omega$ SST model predicts plane-strain turbulence. All flows in the present study have a strain rate tensor which results in at least one zero eigenvalue of $b_{ij}$, for the linear eddy viscosity approximation ($b_{ij} = -\nu_t/k S_{ij}$), and therefore plane-strain turbulence is predicted for all flows by the $k$-$\omega$ SST model. 

As discussed in Section~\ref{sec:realizability_informed_training}, realizability-informed learning function does not guarantee a fully realizable prediction. Rather, realizability-informed learning encodes a preference for realizable predictions into the model predictions. This avoids the need for ad-hoc post-processing of the predicted anisotropy tensor, while also encoding a physics-based (learning) bias into the TBNN. Strict realizability can be further enforced by post-processing any non-realizable predictions by the TBNN, as the anisotropy tensor can still be accessed and assessed for realizability in the proposed framework (albeit, with an evolving linear component).

An important distinction between the dimensional and non-dimensional anisotropy tensor can also be drawn from the results in this section. In turbulence modelling, it is often the non-dimensional anisotropy tensor $b_{ij}$ that is thought to be of interest --- indeed, it is possible to non-dimensionalize the closure problem (e.g., the generalized eddy viscosity model proposed by~\citet{Pope1975}). However, in the present study, we show that even with physically realizable DNS data, dimensionalizing the anisotropy tensor via $a_{ij}=2kb_{ij}$ provides a distinct learning target and loss landscape. The realizability-informed loss function does not in principle compete against a simple error-based loss function in terms of predicting the \textit{non-dimensional} $b_{ij}$ --- non-realizable predictions will also have high error. However, setting the \textit{dimensional} anisotropy tensor as the target creates a tradeoff between these two objectives, as demonstrated by the results in Table~\ref{tbl:realizability_test}. Since predicting the dimensional anisotropy tensor is favourable for the reasons outlined in Section~\ref{sec:realizability_informed_training},further investigation is required to determine the exact source of this tradeoff. This future research area is particularly relevant for learning anisotropy mappings from data, a major area of focus for improving RANS via ML.

\section{Conclusion}\label{sec:conclusion_ri}
The objectives of this study were to propose a physics-informed loss function for training TBNN anisotropy mappings, a new TBNN architecture, and a new injection framework that accommodates implicit treatment of the linear anisotropy component within a TBNN-type architecture. This framework addresses an issue related to the stability of injected TBNN predictions, an issue that has been reported by \citet{Kaandorp2020}, and consistent with our own experience. This framework also addresses the issue of realizability within the context of predicting the anisotropy tensor from RANS input features.

The results here indicate that the proposed model architecture generalizes well to new flow configurations, and the predicted anisotropy tensor can be injected in a highly stable manner. During the injection procedure in the present investigation, the CFD solver remained stable, even when testing erratic model predictions. While this finding corroborates findings from others that frameworks which leverage implicit treatment of the linear anisotropy component via an eddy viscosity are numerically stable~\citep{Wu2019,Brener2021,Liu2021}, a novelty in the present investigation is the avoidance of using an optimal eddy viscosity. With the proposed decomposition of $a_{ij}$ (Section~\ref{sec:injection_ri}), a simpler (and more straightforward) baseline $k$-$\omega$ SST eddy viscosity is used. The core modification that allows a well-conditioned solution in this case is the use of $S^\theta_{ij}$ within $\hat{T}^{(1)}_{ij}$ in the TBNN training process. As a result, all corrections induced by the TBNN are contained in an explicit term in the momentum equation. Future work includes investigating how new-flow generalization can be improved by a blending factor that multiplies this explicit term, thereby allowing the correction to be turned off. For example, a statistics-based scalar or a separate machine learning model could be used to predict a blending factor. This blending factor could reduce or eliminates corrections when a test data point has significant departure from the training dataset, and erroneous predictions are likely.

While the realizability-informed loss function does not strictly guarantee physical realizability of the predictions, the results in Section~\ref{sec:realizability_results} indicate that the model retains a realizability bias when generalizing. The realizability-informed loss function is not only applicable to the framework and architecture proposed in this study---it could be used anytime an anisotropy mapping is generated via machine learning. The use of the realizability-informed loss function in the present investigation promoted better realizability of predictions by a TBNN, but we fully expect the bias induced by this loss function to also be beneficial for tensor basis random forests (e.g., \citet{Kaandorp2020}), or non-tensor basis frameworks (e.g., \citet{Wu2018}). Since high-quality DNS anisotropy tensor data is realizable, the realizability penalty can be viewed as an additional boost to the loss function gradient towards the true value, when a non-realizable prediction is made. Future work will investigate how the realizability-informed loss function performs with training data generated by LES, which is not guaranteed to be realizable.

The KCNN used in the proposed framework to correct $k$ could also be replaced by coupling the $k$ equation to the momentum equation, as exists in the original turbulence model, and the TBNN approach originally proposed by \citet{Ling2016}. At training time, $k^\theta$ is used to dimensionalize $b_{ij}$ in the present study, since $k$ is corrected via the KCNN. This direct correction produced satisfactory results in the present work, but it is possible that generalization could be further enhanced by the re-coupling the $k$ equation with an updated closure term. This enhanced generalization would result from the fact that a physics-based coupled equation system is used to correct $k$, rather than a simple multiplicative corrector (the KCNN in the present investigation). However, this coupling of an additional partial differential equation introduces the possibility for instability, an issue which is common in machine learning anisotropy modelling. Future work will investigate the merits of this route.

Ultimately, this investigation demonstrates that with sufficient modifications, TBNN-type an\-iso\-tropy mappings can be injected in a stable and well-conditioned manner. Further, with appropriate physics-based loss function penalties, the mapping can be sensitized to more physically informative targets than mean-squared error. Moreover, we provided a preliminary investigation of the utility of KANs for turbulence closure modeling. While TBKAN did not outperform TBNN in this context, they nevertheless demonstrated the potential for capturing the complex relationships in the anisotropy tensor (at least in the flat plate case), suggesting that future research work should be conducted on the use of KANs for turbulent closure modeling applications. While industrial use of machine learning-based anisotropy mappings is currently not widespread, the continual development of techniques which increase the practicality of training and injecting model predictions will help accelerate more widespread use. Machine learning-augmented turbulence closure modelling is an aid that the turbulence modelling community can use to help bridge the current computational gap between RANS and widespread use of LES~\citep{Witherden2017}.

\backsection[Acknowledgements]{We sincerely thank the anonymous reviewers for their careful review and helpful suggestions for the manuscript.}

\backsection[Funding]{This work was supported by the Natural Sciences and Engineering Research Council of Canada (NSERC) Postgraduate Scholarship program. Computational resources for this work were supported by the Tyler Lewis Clean Energy Research Foundation.}

\backsection[Declaration of interests]{The authors report no conflict of interest.}

\backsection[Data availability statement]{All code and data that support the results of this study are publicly available. The code can be found in the following repositories: http://doi.org/[doi], http://doi.org/[doi]. The dataset used in this work is available at http://doi.org/[doi]. }

\backsection[Author ORCIDs]{R. McConkey, https://orcid.org/0000-0003-0674-1849; N. Kalia https://orcid.org/0009-0005-5032-6660; E. Yee, https://orcid.org/0000-0002-6413-4329; F.S. Lien https://orcid.org/0000-0003-4301-2610}

\backsection[Author contributions]{RM: conceptualization, methodology, software, validation, investigation,  data curation, writing (original draft, review, and editing), visualization. NK (KAN related work): conceptualization, methodology, software, validation, investigation, writing (original draft, review, and editing), visualization. EY: conceptualization, methodology writing (review and editing), supervision, project administration, funding acquisition. FSL: conceptualization, methodology, writing (review and editing), supervision, project administration, funding acquisition. }

\newpage
\appendix
\section{Integrity Basis Input Features}
\label{ap:symbolic_results}

Wu et al.'s integrity basis is derived from four gradient tensors: $\hat{S}$, $\hat{R}$, $\hat{A}_p$, and $\hat{A}_k$ \citep{Wu2018}. These tensors are calculated as follows:

\begin{align}
    \hat{S}_{ij}&= C_{S} \frac{1}{2}\left(\frac{\partial U_i}{\partial x_j} + \frac{\partial U_j}{\partial x_i}\right) \ , \\
    \hat{R}_{ij} &= C_{R}\frac{1}{2}\left(\frac{\partial U_i}{\partial x_j} - \frac{\partial U_j}{\partial x_i}\right) \ , \\
    \hat{A}_p &= C_{A_p} \epsilon_{ijl} \frac{\partial p}{\partial x_l} \ , \\
    \hat{A}_k&= C_{A_k}\epsilon_{ijl} \frac{\partial k}{\partial x_l} \ , \\
\end{align}
where $C_S$, $C_R$, $C_{A_p}$, and $C_{A_k}$ are scalars which non-dimensionalize their corresponding gradient tensors, and $\epsilon_{ijl}$ is the Levi-Civita symbol. For example, \citet{Ling2016} chose $C_{S}=C_{R}=k/\varepsilon$, so that $\hat{S}$ is dimensionless.
 
Without loss of generality, the scalars $p$ and $k$ in the above can be swapped out for other scalars, such as $\varepsilon$, or $\omega$, thereby replacing $\hat{A}_{p,ij}$ and $\hat{A}_{k,ij}$ with $\hat{C_{ij}}$ and $\hat{D}_{ij}$:
\begin{align}
    \hat{C}_{ij} &= C_{C} \epsilon_{ijl} \frac{\partial s_1}{\partial x_l} \ , \\
    \hat{D}_{ij}&= C_{D}\epsilon_{ijl} \frac{\partial s_2}{\partial x_l} \ , \\
\end{align}
where $s_1$ and $s_2$ are two scalar fields. For example, in the present study, $s_1=\omega$, and $s_2=k$. In three dimensional cartesian coordinates, the strain rate, rotation rate, and antisymmetric tensors associated with the two scalar gradients are:
\begin{align}
    \hat{S}_{ij} &=\frac{C_S}{2}\begin{bmatrix}
    2\dfrac{\strut\partial U_1}{\strut\partial x_1} & \dfrac{\strut\partial U_1}{\strut\partial x_2} + \dfrac{\strut\partial U_2}{\strut\partial x_1} & \dfrac{\strut\partial U_1}{\strut\partial x_3} + \dfrac{\strut\partial U_3}{\strut\partial x_1}\\
    \dfrac{\strut\partial U_2}{\strut\partial x_1}+\dfrac{\strut\partial U_1}{\strut\partial x_2} & 2\dfrac{\strut\partial U_2}{\strut\partial x_2} & \dfrac{\strut\partial U_2}{\strut\partial x_3} + \dfrac{\strut\partial U_3}{\strut\partial x_2}\\
    \dfrac{\strut\partial U_3}{\strut\partial x_1}+\dfrac{\strut\partial U_1}{\strut\partial x_3}  & \dfrac{\strut\partial U_3}{\strut\partial x_2} +\dfrac{\strut\partial U_2}{\strut\partial x_3} & 2\dfrac{\strut\partial U_3}{\strut\partial x_3}
    \end{bmatrix} \ , \\
    \hat{R}_{ij}&= \frac{C_R}{2}\begin{bmatrix}
    0 & \dfrac{\strut\partial U_1}{\strut\partial x_2} - \dfrac{\strut\partial U_2}{\strut\partial x_1} & \dfrac{\strut\partial U_1}{\strut\partial x_3} - \dfrac{\strut\partial U_3}{\strut\partial x_1}\\
    \dfrac{\strut\partial U_2}{\strut\partial x_1}-\dfrac{\strut\partial U_1}{\strut\partial x_2} & 0 & \dfrac{\strut\partial U_2}{\strut\partial x_3} - \dfrac{\strut\partial U_3}{\strut\partial x_2}\\
    \dfrac{\strut\partial U_3}{\strut\partial x_1}-\dfrac{\strut\partial U_1}{\strut\partial x_3}  & \dfrac{\strut\partial U_3}{\strut\partial x_2} -\dfrac{\strut\partial U_2}{\strut\partial x_3} & 0
    \end{bmatrix}  \ , \\
    \hat{C}_{ij} &= C_C \begin{bmatrix}
    0 & -\dfrac{\strut\partial s_1}{\strut\partial x_3} & \dfrac{\strut\partial s_1}{\strut\partial x_2}\\
    \dfrac{\strut\partial s_1}{\strut\partial x_3} & 0 & -\dfrac{\strut\partial s_1}{\strut\partial x_1}\\
    -\dfrac{\partial s_1}{\strut\partial x_2} & \dfrac{\strut\partial s_1}{\strut\partial x_1} & 0
    \end{bmatrix} \ , \\
   \hat{D}_{ij} &= C_D \begin{bmatrix}
    0 & -\dfrac{\strut\partial s_2}{\strut\partial x_3} & \dfrac{\strut\partial s_2}{\strut\partial x_2}\\
    \dfrac{\strut\partial s_2}{\strut\partial x_3} & 0 & -\dfrac{\strut\partial s_2}{\strut\partial x_1}\\
    -\dfrac{\strut\partial s_2}{\strut\partial x_2} & \dfrac{\strut\partial s_2}{\strut\partial x_1} & 0
    \end{bmatrix} \ .
\end{align}

Under some conditions, input features derived from invariants of the minimal integrity basis derived by \citet{Wu2018} can vanish. These conditions occur when there are zero-gradients in the flow, as all tensors in Wu's et al.'s integrity basis are derived from gradient-based tensors.

Here, we consider two cases:
\begin{itemize}
    \item [\textbf{(I)}] \textbf{Two-dimensional flow}.

    The zero pressure gradient boundary layer and periodic hills cases in the present study fall into this category. With the coordinate system defined as it was in Section~\ref{sec:periodic_hills}, the velocity vector is $U_j = (U_1, U_2, 0)$, and the gradient tensors take the following form in three dimensional space:
    
\begin{align}
    \hat{S}_{ij} &=\frac{C_S}{2}\begin{bmatrix}
    2\dfrac{\strut\partial U_1}{\strut\partial x_1} & \dfrac{\strut\partial U_1}{\strut\partial x_2} + \dfrac{\strut\partial U_2}{\strut\partial x_1} & 0\\
    \dfrac{\strut\partial U_2}{\strut\partial x_1}+\dfrac{\strut\partial U_1}{\strut\partial x_2} & 2\dfrac{\strut\partial U_2}{\strut\partial x_2} & 0\\
    0 & 0 & 0
    \end{bmatrix} \ , \\
    \hat{R}_{ij}&= \frac{C_R}{2}\begin{bmatrix}
    0 & \dfrac{\strut\partial U_1}{\strut\partial x_2} - \dfrac{\strut\partial U_2}{\strut\partial x_1} & 0\\
    \dfrac{\strut\partial U_2}{\strut\partial x_1}-\dfrac{\strut\partial U_1}{\strut\partial x_2} & 0 & 0\\
    0 & 0 & 0
    \end{bmatrix} \ , \\
    \hat{C}_{ij} &= C_C \begin{bmatrix}
    0 & 0 & \dfrac{\strut\partial s_1}{\strut\partial x_2}\\
    0 & 0 & -\dfrac{\strut\partial s_1}{\strut\partial x_1}\\
    -\dfrac{\partial s_1}{\strut\partial x_2} & \dfrac{\strut\partial s_1}{\strut\partial x_1} & 0
    \end{bmatrix} \ , \\
   \hat{D}_{ij} &= C_D \begin{bmatrix}
    0 & 0 & \dfrac{\strut\partial s_2}{\strut\partial x_2}\\
    0 & 0 & -\dfrac{\strut\partial s_2}{\strut\partial x_1}\\
    -\dfrac{\strut\partial s_2}{\strut\partial x_2} & \dfrac{\strut\partial s_2}{\strut\partial x_1} & 0  \ .
    \end{bmatrix} 
\end{align}
    
    \item [\textbf{(II)}] \textbf{Three-dimensional flow with zero gradients in one direction.}.

    The square duct case in the present study falls into this category. With the coordinate system defined as it was in Section~\ref{sec:square_duct}, the velocity vector is $U_j = (U_1, U_2, U_3)$. All gradients in the $x_1$ direction are zero: $\partial()/\partial x_1 = 0$. In this case, the gradient tensors take the following form in three dimensional space:
    \begin{align}
    \hat{S}_{ij} &=\frac{C_S}{2}\begin{bmatrix}
    0 & \dfrac{\strut\partial U_1}{\strut\partial x_2}  & \dfrac{\strut\partial U_1}{\strut\partial x_3} \\
    \dfrac{\strut\partial U_1}{\strut\partial x_2} & 2\dfrac{\strut\partial U_2}{\strut\partial x_2} & \dfrac{\strut\partial U_2}{\strut\partial x_3} + \dfrac{\strut\partial U_3}{\strut\partial x_2}\\
    \dfrac{\strut\partial U_1}{\strut\partial x_3}  & \dfrac{\strut\partial U_3}{\strut\partial x_2} +\dfrac{\strut\partial U_2}{\strut\partial x_3} & 2\dfrac{\strut\partial U_3}{\strut\partial x_3}
    \end{bmatrix} \ , \\
    \hat{R}_{ij}&= \frac{C_R}{2}\begin{bmatrix}
    0 & \dfrac{\strut\partial U_1}{\strut\partial x_2}& \dfrac{\strut\partial U_1}{\strut\partial x_3} \\
    \dfrac{\strut\partial U_1}{\strut\partial x_2} & 0 & \dfrac{\strut\partial U_2}{\strut\partial x_3} - \dfrac{\strut\partial U_3}{\strut\partial x_2}\\
    \dfrac{\strut\partial U_1}{\strut\partial x_3}  & \dfrac{\strut\partial U_3}{\strut\partial x_2} -\dfrac{\strut\partial U_2}{\strut\partial x_3} & 0
    \end{bmatrix}  \ , \\
    \hat{C}_{ij} &= C_C \begin{bmatrix}
    0 & -\dfrac{\strut\partial s_1}{\strut\partial x_3} & \dfrac{\strut\partial s_1}{\strut\partial x_2}\\
    \dfrac{\strut\partial s_1}{\strut\partial x_3} & 0 & 0\\
    -\dfrac{\partial s_1}{\strut\partial x_2} & 0 & 0
    \end{bmatrix} \ , \\
   \hat{D}_{ij} &= C_D \begin{bmatrix}
    0 & -\dfrac{\strut\partial s_2}{\strut\partial x_3} & \dfrac{\strut\partial s_2}{\strut\partial x_2}\\
    \dfrac{\strut\partial s_2}{\strut\partial x_3} & 0 & 0\\
    -\dfrac{\strut\partial s_2}{\strut\partial x_2} & 0 & 0  \ .
    \end{bmatrix} 
\end{align}

\end{itemize}

Cases \textbf{(I)} and \textbf{(II)} were analyzed using sympy~\citep{sympy} to determine which integrity basis tensor invariants are non-zero, and therefore suitable as potential input features. The first and second invariants of a rank two tensor are scalar functions, defined by:

\begin{align}
    I_1(A_{ij}) &= A_{ii}\\
    I_2(A_{ij})&= \frac{1}{2}\left(\left(A_{ii}\right)^2 - A_{ij}A_{ji}\right)
\end{align}
The third invariant, $I_3=\textbf{det}(A_{ij})$ is zero for all of the integrity basis tensors, since they are either antisymmetric, or symmetric and zero trace. Table~\ref{tbl:invariants} shows the results of this analysis.

\renewcommand{\arraystretch}{1.5}

\begin{longtable}{|p{1cm}|p{4cm}|p{1cm}|p{1cm}|p{1cm}|p{1cm}|}
%\centering
\caption{Non-zero invariants of the minimal integrity basis tensor formed by $\hat{S}_{ij}$, $\hat{R}_{ij}$, $\hat{C}_{ij}$, and $\hat{D}_{ij}$.}\label{tbl:invariants}\\

\hline \multicolumn{1}{|c|}{\textbf{Tensor}} & \multicolumn{1}{c|}{\textbf{Expression}} & \multicolumn{1}{c|}{\textbf{(I)}$I_1\neq 0$ } & \multicolumn{1}{c|}{\textbf{(I)}$I_2\neq 0$ } & \multicolumn{1}{c|}{\textbf{(II)}$I_1\neq 0$}& \multicolumn{1}{c|}{\textbf{(II)}$I_2\neq 0$ } \\ \hline
\endfirsthead

\multicolumn{6}{c}%
        {{ Table \thetable\ (continued)}} \\
\hline \multicolumn{1}{|c|}{\textbf{Tensor}} & \multicolumn{1}{c|}{\textbf{Expression}} & \multicolumn{1}{c|}{\textbf{(I)}$I_1\neq 0$ } & \multicolumn{1}{c|}{\textbf{(I)}$I_2\neq 0$ } & \multicolumn{1}{c|}{\textbf{(II)}$I_1\neq 0$}& \multicolumn{1}{c|}{\textbf{(II)}$I_2\neq 0$ } \\ \hline
\endhead

\hline
\endlastfoot

\hline
\multicolumn{6}{c}%
        {{(continued on next page)}}
\endfoot

%\hline test & test & test &test & test & test\\
%\hline \endhead

$B^{(1)}_{ij}$   & $\hat{S}_{ik}\hat{S}_{kj}$       & $\checkmark$& $\checkmark$& $\checkmark$ &  $\checkmark$            \\
$B^{(2)}_{ij}$   & $\hat{S}_{ik}\hat{S}_{kl}\hat{S}_{lj}$       & ---       & $\checkmark$& ---       & $\checkmark$       \\
$B^{(3)}_{ij}$   & $\hat{R}_{ik}\hat{R}_{kj}$       & $\checkmark$& $\checkmark$& $\checkmark$  & $\checkmark$       \\
$B^{(4)}_{ij}$   & $\hat{C}_{ik}\hat{C}_{kj}$  & $\checkmark$& $\checkmark$& $\checkmark$     & $\checkmark$              \\
$B^{(5)}_{ij}$   & $\hat{D}_{ik}\hat{D}_{kj}$     & $\checkmark$& $\checkmark$& $\checkmark$     & $\checkmark$        \\
$B^{(6)}_{ij}$   & $\hat{R}_{ik}\hat{R}_{kl}\hat{S}_{lj}$      &  ---      & $\checkmark$& ---       & $\checkmark$        \\
$B^{(7)}_{ij}$   & $\hat{R}_{ik}\hat{R}_{kl}\hat{S}_{lm}\hat{S}_{mj}$    & $\checkmark$& $\checkmark$& $\checkmark$   & $\checkmark$\\
$B^{(8)}_{ij}$   & $\hat{R}_{ik}\hat{R}_{kl}\hat{S}_{lm} \hat{R}_{mn} \hat{S}_{no} \hat{S}_{oj} $  & --- & $\checkmark$& ---  & $\checkmark$    \\
$B^{(9)}_{ij}$   & $\hat{C}_{ik}\hat{C}_{kl}\hat{S}_{lj}$         & $\checkmark$& ---       & ---     & $\checkmark$     \\
$B^{(10)}_{ij}$& $\hat{C}_{ik}\hat{C}_{kl}\hat{S}_{lm}\hat{S}_{mj} $  & $\checkmark$& --- & $\checkmark$     & $\checkmark$         \\
$B^{(11)}_{ij}$& $\hat{C}_{ik}\hat{C}_{kl}\hat{S}_{lm}\hat{C}_{mn}\hat{S}_{no}\hat{S}_{oj}$ &  --- & --- & $\checkmark$    & $\checkmark$     \\
$B^{(12)}_{ij}$& $\hat{D}_{ik}\hat{D}_{kl}\hat{S}_{lj}$ & $\checkmark$& ---       & ---     & $\checkmark$         \\
$B^{(13)}_{ij}$& $\hat{D}_{ik}\hat{D}_{kl}\hat{S}_{lm}\hat{S}_{mj}$  & $\checkmark$& ---  & $\checkmark$     & $\checkmark$         \\
$B^{(14)}_{ij}$& $\hat{D}_{ik}\hat{D}_{kl}\hat{S}_{lm}\hat{D}_{mn}\hat{S}_{no}\hat{S}_{oj}$  & --- &  --- & $\checkmark$   & $\checkmark$   \\
$B^{(15)}_{ij}$& $\hat{R}_{ik}\hat{C}_{kj}$ & ---       &  ---      & $\checkmark$   & $\checkmark$         \\
$B^{(16)}_{ij}$& $\hat{C}_{ik}\hat{D}_{kj}$ & $\checkmark$& $\checkmark$& $\checkmark$     &  $\checkmark$     \\
$B^{(17)}_{ij}$& $\hat{R}_{ik}\hat{D}_{kj}$      & ---       & ---       & $\checkmark$   & $\checkmark$         \\
$B^{(18)}_{ij}$& $\hat{R}_{ik}\hat{C}_{kl}\hat{S}_{lj}$  & ---       & ---       & ---         & $\checkmark$         \\
$B^{(19)}_{ij}$& $\hat{R}_{ik}\hat{C}_{kl}\hat{S}_{lm}\hat{S}_{mj}$    & ---    & ---   & $\checkmark$    & $\checkmark$         \\
$B^{(20)}_{ij}$& $\hat{R}_{ik}\hat{R}_{kl}\hat{C}_{lm}\hat{S}_{mj} $         & ---       & ---       & ---  & $\checkmark$         \\
$B^{(21)}_{ij}$& $\hat{C}_{ik}\hat{C}_{kl}\hat{R}_{lm}\hat{S}_{mj}  $  & $\checkmark$& ---& $\checkmark$     & $\checkmark$      \\
$B^{(22)}_{ij}$& $\hat{R}_{ik}\hat{R}_{kl}\hat{C}_{lm}\hat{S}_{mn}\hat{S}_{nj} $       & ---       & --- & --- & $\checkmark$         \\
$B^{(23)}_{ij}$& $\hat{C}_{ik}\hat{C}_{kl} \hat{R}_{lm} \hat{S}_{mn} \hat{S}_{nj}$      & ---       & --- & --- & $\checkmark$         \\
$B^{(24)}_{ij}$& $\hat{R}_{ik}\hat{R}_{kl}\hat{S}_{lm}\hat{C}_{mn}\hat{S}_{no}\hat{S}_{oj}$  & ---       & ---  & ---  & $\checkmark$         \\
$B^{(25)}_{ij}$& $\hat{C}_{ik}\hat{C}_{kl}\hat{S}_{lm}\hat{R}_{mn}\hat{S}_{no}\hat{S}_{oj}$  & $\checkmark$& ---  & $\checkmark$ & $\checkmark$ \\
$B^{(26)}_{ij}$& $\hat{R}_{ik}\hat{D}_{kl}\hat{S}_{lj}$  &  ---      & ---       & ---         & $\checkmark$  \\
$B^{(27)}_{ij}$& $\hat{R}_{ik}\hat{D}_{kl}\hat{S}_{lm}\hat{S}_{mj} $         & ---  & --- & $\checkmark$  & $\checkmark$         \\
$B^{(28)}_{ij}$& $\hat{R}_{ik} \hat{R}_{kl} \hat{D}_{lm} \hat{S}_{mj}$   & ---       & ---       & --- & $\checkmark$     \\
$B^{(29)}_{ij}$& $\hat{D}_{ik} \hat{D}_{kl} \hat{R}_{lm} \hat{S}_{mj}$ & $\checkmark$& --- & $\checkmark$ & $\checkmark$          \\
$B^{(30)}_{ij}$& $\hat{R}_{ik} \hat{R}_{kl} \hat{D}_{lm} \hat{S}_{mn} \hat{S}_{nj}$ & ---       & ---   & --- & $\checkmark$         \\
$B^{(31)}_{ij}$& $\hat{D}_{ik} \hat{D}_{kl} \hat{R}_{lm} \hat{S}_{mn} \hat{S}_{nj}$ & ---       & ---       & --- & $\checkmark$         \\
$B^{(32)}_{ij}$& $\hat{R}_{ik} \hat{R}_{kl}\hat{S}_{lm} \hat{D}_{mn} \hat{S}_{no} \hat{S}_{oj}$ & ---    & ---  & --- & $\checkmark$         \\
$B^{(33)}_{ij}$& $\hat{D}_{ik} \hat{D}_{kl} \hat{S}_{lm} \hat{R}_{mn} \hat{S}_{no} \hat{S}_{oj}$ & $\checkmark$& ---& $\checkmark$ & $\checkmark$\\
$B^{(34)}_{ij}$& $\hat{C}_{ik} \hat{D}_{kl} \hat{S}_{lj}$         & $\checkmark$& ---       & ---     & $\checkmark$         \\
$B^{(35)}_{ij}$& $\hat{C}_{ik} \hat{D}_{kl} \hat{S}_{lm} \hat{S}_{mj}$  & $\checkmark$& --- & $\checkmark$ & $\checkmark$         \\
$B^{(36)}_{ij}$& $\hat{C}_{ik} \hat{C}_{kl} \hat{D}_{lm} \hat{S}_{mj}$       & --- & --- & $\checkmark$ & $\checkmark$         \\
$B^{(37)}_{ij}$& $\hat{D}_{ik} \hat{D}_{kl} \hat{C}_{lm} \hat{S}_{mj}$       & ---    & ---  & $\checkmark$  & $\checkmark$         \\
$B^{(38)}_{ij}$& $\hat{C}_{ik} \hat{C}_{kl} \hat{D}_{lm} \hat{S}_{mn} \hat{S}_{nj}$  & ---       & ---       & --- & $\checkmark$         \\
$B^{(39)}_{ij}$& $\hat{D}_{ik} \hat{D}_{kl} \hat{C}_{lm} \hat{S}_{mn} \hat{S}_{nj}$   & ---       & --- & ---  & $\checkmark$         \\
$B^{(40)}_{ij}$& $\hat{C}_{ik} \hat{C}_{kl} \hat{S}_{lm} \hat{D}_{mn} \hat{S}_{no} \hat{S}_{oj}$ & ---  & --- & $\checkmark$ & $\checkmark$     \\
$B^{(41)}_{ij}$& $\hat{D}_{ik} \hat{D}_{kl} \hat{S}_{lm}\hat{C}_{mn}\hat{S}_{no} \hat{S}_{oj}$   & --- & --- & $\checkmark$ & $\checkmark$    \\
$B^{(42)}_{ij}$& $\hat{R}_{ik}\hat{C}_{kl} \hat{D}_{lj}$         & $\checkmark$& ---       & ---     & $\checkmark$         \\
$B^{(43)}_{ij}$& $\hat{R}_{ik} \hat{C}_{kl} \hat{D}_{lm} \hat{S}_{mj}$  & $\checkmark$& ---   & $\checkmark$ & $\checkmark$         \\
$B^{(44)}_{ij}$& $\hat{R}_{ik} \hat{D}_{kl} \hat{C}_{lm} \hat{S}_{mj}$     & $\checkmark$& ---  & $\checkmark$ & $\checkmark$      \\
$B^{(45)}_{ij}$& $\hat{R}_{ik} \hat{C}_{kl} \hat{D}_{lm} \hat{S}_{mn} \hat{S}_{nj}$    & $\checkmark$& ---  & --- & $\checkmark$         \\
$B^{(46)}_{ij}$& $\hat{R}_{ik} \hat{D}_{kl} \hat{C}_{lm} \hat{S}_{mn} \hat{S}_{nj}$   & $\checkmark$& --- & ---     & $\checkmark$      \\
$B^{(47)}_{ij}$& $\hat{R}_{ik} \hat{C}_{kl} \hat{S}_{lm} \hat{D}_{mn} \hat{S}_{no} \hat{S}_{oj}$   & ---  & --- & --- & $\checkmark$         \\ \hline
\end{longtable}

Source code which supports the analysis in this appendix and enables further investigation is available on Github~\citep{integritybasisgithub}.

\bibliographystyle{jfm}
\bibliography{references}

\end{document}